\documentclass[showpacs, preprint,preprintnumbers,showpacs,showkeys,superscriptaddress,amsmath,amssymb,nofootinbib]{revtex4-1}
\usepackage{amsmath}
\usepackage{latexsym}
\usepackage{pstricks,pst-coil}
\usepackage{pst-all}
\usepackage{comment}

\usepackage{color}
\usepackage{latexsym}
\usepackage{amsmath}
\usepackage{amssymb}
\usepackage{eufrak}
\usepackage{euscript}
\usepackage[latin1]{inputenc}
\usepackage{pstricks}
\usepackage{pst-all}
\usepackage{epstopdf}
\usepackage{epsfig}
\usepackage{graphics}
\usepackage{graphicx}
\usepackage{picture}

\newcommand{\be}{\begin{equation}}
\newcommand{\ee}{\end{equation}}
\newcommand{\ba}{\begin{eqnarray}}
\newcommand{\ea}{\end{eqnarray}}

\newcommand{\bea}{\begin{eqnarray}}
\newcommand{\eea}{\end{eqnarray}}

\begin{document}

\title{\Large{
Majorana Fermion Dark Matter \\
in Minimally Extended Left-Right Symmetric Model
}}

\author{M. J. Neves}\email{mneves@ua.edu}
\affiliation{Department of Physics and Astronomy, University of Alabama, Tuscaloosa, AL 35487, USA}
\affiliation{Departamento de F\'isica, Universidade Federal Rural do Rio de Janeiro,
BR 465-07, 23890-971, Serop\'edica, RJ, Brazil}
\author{Nobuchika Okada}\email{okadan@ua.edu}
\affiliation{Department of Physics and Astronomy, University of Alabama, Tuscaloosa, AL 35487, USA}
\author{Satomi Okada}\email{satomi.okada@ua.edu}
\affiliation{Department of Physics and Astronomy, University of Alabama, Tuscaloosa, AL 35487, USA}




\begin{abstract}
\noindent
We present a minimal extension of the left-right symmetric model 
   based on the gauge group $SU(3)_{c} \times SU(2)_{L} \times SU(2)_{R} \times U(1)_{B-L} \times U(1)_{X}$,
   in which a vector-like fermion pair ($\zeta_L$ and $\zeta_R$) charged 
   under the $U(1)_{B-L} \times U(1)_X$ symmetry is introduced. 
Associated with the symmetry breaking of the gauge group $SU(2)_{R} \times U(1)_{B-L} \times U(1)_{X}$ 
   down to the Standard Model (SM) hypercharge $U(1)_Y$,  
   Majorana masses for $\zeta_{L, R}$ are generated and the lightest mass eigenstate 
   plays a role of the dark matter (DM) in our universe by its communication with the SM particles 
   through a new neutral gauge boson ``$X$". 
We consider various phenomenological constraints of this DM scenario, 
   such as the observed DM relic density, 
   the LHC Run-2 constraints from the search for a narrow resonance, 
   and the perturbativity of the gauge couplings below the Planck scale. 
Combining all constraints, we identify the allowed parameter region which turns out to be very narrow. 
A significant portion of the currently allowed parameter region will be tested 
   by the High-Luminosity LHC experiments.  
\end{abstract}

%

\maketitle

\section{Introduction}
\label{sec:1}

The left-right symmetric model (LRSM) is one of well-motivated models beyond the Standard Model (SM), 
  which was introduced for understanding the origin of the parity violation in the SM 
  \cite{PatiPRD1974,MohapatraPRD1975,SenjanoviPRD1975}.
The model is based on the gauge group $SU(3)_c \times SU(2)_{L} \times SU(2)_{R} \times U(1)_{B-L}$.
The leptonic $SU(2)_{R}$ doublet includes the right-handed neutrinos (RHNs), 
  and the spontaneous symmetry breaking of $SU(2)_{R} \times U(1)_{B-L}$ to the SM $U(1)_Y$ 
  generates Majorana masses for the RHNs.   
With the subsequent electroweak symmetry breaking, tiny SM neutrino masses are naturally generated 
  by the type-I seesaw mechanism \cite{seesaw1, seesaw2, seesaw3, seesaw4, seesaw5}. 
New charged and neutral gauge bosons, $W_{R}$ and $Z_{R}$, predicted by the model 
  have been searched for by the Large Hadron Collider (LHC) experiments \cite{CMS, ATLAS,miha}.

Although the LRSM is very interesting, a DM candidate is missing in its minimal version. 
Simple extensions of the LRSM to incorporate a fermion or scaler DM candidate 
  have been proposed in Refs.~\cite{olive, heeck, amitabha}, 
  and then their DM phenomenologies have been investigated in detail \cite{heeck2,hooper,patra2,patra3},    
  where the DM interactions with the SM particles through $W_R$ and $Z_R$ play a central role. 
In another approach, the LRSM can be minimally extended to incorporate a new $U(1)_X$ gauge group 
  and a Dirac fermion DM which is singlet under the SM gauge group \cite{MJNevesHelayelMohapatraOkada2018} 
  (see also Ref.~\cite{MJNeves2018}). 
In this scenario, the DM particle communicates with the SM particles through a massive gauge boson ($X$), 
  which arises as a linear combination of the $SU(2)_R$, $U(1)_{B-L}$ and $U(1)_X$ gauge bosons
  after the symmetry breaking of $SU(2)_{R} \times U(1)_{B-L} \times U(1)_{X}$ down to the SM $U(1)_Y$.   
This class of DM models is called ``$Z^\prime$-portal DM scenario'' 
   (for a review, see Ref.~\cite{SatomiAHEP2018} and references therein).

In this paper, we consider a Majorana fermion DM in the context of a minimal extension 
  of the LRSM with a new $U(1)_X$ gauge symmetry, which is based on the gauge group 
  $SU(3)_c \times SU(2)_{L} \times SU(2)_{R} \times U(1)_{B-L} \times U(1)_{X}$.  
 As mentioned above, this minimal $U(1)_X$ extension has been proposed 
   in Ref.~\cite{MJNevesHelayelMohapatraOkada2018} 
   to incorporate a Dirac fermion DM, 
   where the $U(1)_X$ symmetry ensures the stability of the Dirac fermion
   and the DM fermion communicates with the SM particles through the massive gauge boson $X$. 
Although the gauge group and the particle content of our model is the same
  as those in Ref.~\cite{MJNevesHelayelMohapatraOkada2018}, 
  we consider in this paper a modification of the $U(1)_{B-L} \times U(1)_X$ charge assignment 
  for the vector-like fermion pair ($\zeta_L$ and $\zeta_R$) 
  by which their Majorana masses are generated by the symmetry breaking, 
  in addition to the Dirac mass.  
As a result, the lightest Majorana mass eigenstate plays a role of the DM in our universe.   
We carefully calculate the gauge boson mass eigenstates after the symmetry breaking 
  of $SU(2)_{R} \times U(1)_{B-L} \times U(1)_{X}$ down to the SM $U(1)_Y$
  to derive the massive gauge boson $X$ couplings with the DM fermion and the SM fermions. 
Errors in the coupling formulas presented in Ref.~\cite{MJNevesHelayelMohapatraOkada2018} 
   will be corrected in this paper. 
We consider various phenomenologies of our DM scenario, such as the observed DM relic density, 
   the LHC Run-2 constraints on the $X$ boson and theoretical consistency, namely, 
   the perturbativity condition on the gauge couplings up to the reduced Planck scale. 
Combining all the constraints, we identify the allowed parameter region, which turns out to be very narrow.

This paper is organized as follows: 
In Sec.~\ref{sec:2}, we present the minimally extended LRSM with a Majorana fermion DM. 
In Sec.~\ref{sec:3}, we discuss the symmetry breaking of the model down to the SM gauge group 
  and derive the gauge boson mass spectrum. 
We also derive the gauge boson interactions with the SM fermions and the DM particle. 
The perturbativity condition of the gauge coupling constants will be investigated in Sec.~\ref{sec:4}. 
In Sec.~\ref{sec:5}, we consider the LHC Run-2 constraints on the $X$ boson mass and its coupling
  with the SM fermions. 
We will see that the allowed mass range of the $X$ boson is very restricted 
  after combining the LHC Run-2 constraints and the perturbativity condition. 
In Sec.~\ref{sec:6}, we analyze the relic density of the Majorana fermion DM and 
  identify the model parameter region to reproduce the observed DM relic density. 
We combine all the constraints to see the allowed parameter region.  
The last section is devoted to conclusions.

\begin{table}[tb]
\centering
\begin{tabular}{|c||c|c|c|c|c|}
\hline
 &  ~$SU(3)_{c}$~ & ~$SU(2)_{L}$~ & ~$SU(2)_{R}$~ & ~$U(1)_{B-L}$~ &  ~$U(1)_X$~ \\
\hline
\hline
$Q_{L}^i$=
$\left(
    \begin{array}{c}
      u_L^i \\
      d_L^i      
    \end{array}
  \right)$  &  $\underline{{\bf 3}}$  &  $\underline{{\bf 2}}$  &  $\underline{{\bf 1}}$  &  $+1/3$ &  $0$  \\
$Q_{R}^i$=
$\left(
    \begin{array}{c}
      u_R^i \\
      d_R^i      
    \end{array}
  \right)$  &  $\underline{{\bf 3}}$ &  $\underline{{\bf 1}}$  & $\underline{{\bf 2}}$ & $+1/3$ &  $0$ \\
\hline
$\Psi_{L}^i$=
$\left(
    \begin{array}{c}
      \nu_L^i \\
      e_L^i      
    \end{array}
  \right)$  & $\underline{{\bf 1}}$  &  $\underline{{\bf 2}}$  &  $\underline{{\bf 1}}$  & $-1$  & $0$  \\
$\Psi_{R}^i$=
$\left(
    \begin{array}{c}
      N_R^i \\
      e_R^i      
    \end{array}
  \right)$ &  $\underline{{\bf 1}}$  &  $\underline{{\bf 1}}$  &  $\underline{{\bf 2}}$  & $-1$ &  $0$ \\
\hline
$\Phi$  & $\underline{{\bf 1}}$  &  $\underline{{\bf 2}}$  &  $\underline{{\bf 2}}$  &  $0$ &  $0$  \\
$\Delta_{L}$  &  $\underline{{\bf 1}}$ & $\underline{{\bf 3}}$  & $\underline{{\bf 1}}$ & $+2$  &$0$   \\
$\Delta_{R}$ &  $\underline{{\bf 1}}$  & $\underline{{\bf 1}}$  & $\underline{{\bf 3}}$ & $+2$  &  $0$   \\
$\phi_{X}$  &  $\underline{{\bf 1}}$  &  $\underline{{\bf 1}}$ & $\underline{{\bf 1}}$  &  $+a$  & $-a$   \\
\hline
$\zeta_{L, R}$ &  $\underline{{\bf 1}}$  & $\underline{{\bf 1}}$ & $\underline{{\bf 1}}$  &  $-a/2$  & $+ a/2$ \\
\hline
\end{tabular}
\caption{
The particle content of our minimally extended LRSM with Majorana fermion DM. 
Along with the $U(1)_X$ gauge symmetry, a new scalar $\phi_X$ and a vector-like pair of the fermions $\zeta_{L, R}$ are introduced. 
All fields in the original LRSM are singlet under the $U(1)_X$. 
$i=1,2,3$ is the generation index, and $a \neq 0$ is a real parameter. 
}
\label{Table:1}
\end{table}

\section{Minimally extended LRSM with Majorana fermion DM} 
\label{sec:2}
As has been first proposed in Ref.~\cite{MJNevesHelayelMohapatraOkada2018}, 
   the minimally extended LRSM is based on the gauge group 
   ${\cal G}_{\rm LRX} \equiv SU(3)_c\times SU(2)_L\times SU(2)_R\times U(1)_{B-L}\times U(1)_X$. 
The introduction of the new gauge symmetry $U(1)_X$ is the key of the extension.   
The particle content of the model is listed in Table~\ref{Table:1}.
Along with the $U(1)_X$ gauge symmetry, a new scalar $\phi_X$ and a vector-like pair of the fermions $\zeta_{L, R}$ are introduced. 
All fields in the original LRSM are singlet under the $U(1)_X$. 
Note that the charge assignment for $\zeta_{L, R}: (-a/2, a/2)$ 
   is crucial to generate Majorana mass terms for $\zeta_{L, R}$, 
   while $(-b/2, b/2)$ with $b \neq a$ is assigned in Ref.~\cite{MJNevesHelayelMohapatraOkada2018}.

The kinetic terms for the fermions are expressed as
\begin{equation}\label{Lfermions}
{\cal L}_{f}=
 \overline{\psi_L^i} i \gamma^\mu D_\mu \psi_L^i
+\overline{\psi_R^i} i \gamma^\mu D_\mu \psi_R^i
+\overline{Q_L^i} i \gamma^\mu D_\mu Q_L^i
+\overline{Q_R^i} i \gamma^\mu D_\mu Q_R^i
+\overline{\zeta_L} i \gamma^\mu D_\mu \zeta_L
+\overline{\zeta_R} i \gamma^\mu D_\mu \zeta_R,
\end{equation}
where the covariant derivative $D_\mu$ (relevant for $SU(2)_L\times SU(2)_R\times U(1)_{B-L}\times U(1)_X$)  
  is given by
\begin{eqnarray}\label{DmuPsiLPsiR}
D_{\mu} = \partial_{\mu}+i g_{L} A_{L\mu}^{a} \frac{\sigma_{L}^{a}}{2}
+i g_{R} A_{R\mu}^{a} \frac{\sigma_{R}^{a}}{2} + i g_{BL} \frac{Q_{BL}}{2} B_{\mu}
+ i g_{X} \frac{Q_X}{2} C_{\mu} ,
\end{eqnarray}
with corresponding gauge couplings, $g_{L}$, $g_{R}$, $g_{BL}$ and $g_{X}$,  
  and gauge bosons, $A_{L\mu}^{\, a}$, $A_{R\mu}^{\, a}$, $B_{\mu}$, and $C_{\mu}$.
We impose the left-right parity symmetry, so that $g_{L}=g_{R} \equiv g$. 
The most general gauge bosons kinetic terms are given by 
\begin{equation}\label{Lgauge}
{\cal L}_{gauge}=-\frac{1}{2} \,\mbox{tr}\left(F_{L\mu\nu}^{\; 2}\right)
-\frac{1}{2} \,\mbox{tr}\left(F_{R\mu\nu}^{\; 2}\right)
-\frac{1}{4} \, B_{\mu\nu}^{\; 2}
-\frac{1}{4} \, C_{\mu\nu}^{\; 2}
-\frac{1}{2} \, \chi_{mix} \, B_{\mu\nu}C^{\mu\nu} ,
\end{equation}
where $F_{L\mu\nu}$, $F_{R\mu\nu}$, $B_{\mu\nu}$ and $C_{\mu\nu}$ are the field-strength tensors
   of $A_{L\mu}^{\, a}$, $A_{R\mu}^{\, a}$, $B_{\mu}$, and $C_{\mu}$, respectively. 
Although the general Lagrangian includes a kinetic mixing between $B_{\mu\nu}$ and $C_{\mu\nu}$, 
   we set the mixing parameter $\chi_{mix}=0$ through out this paper, for simplicity.

In the minimal LRSM, the Higgs potential for $\Phi$ and $\Delta_{L,R}$ and the symmetry breaking 
  have been investigated in detail \cite{SenjanovicPRD, kayser, baren, kiers, dev, miha0, miha1,goran2, dev2019}. 
We extend the Higgs potential by adding $\phi_{X}$ as follows:  
\begin{eqnarray}\label{LHiggs}
V\left(\Phi, \Delta_{L,R},\phi_{X}\right) &= & - \mu_1^2 \: {\rm Tr} (\Phi^{\dag} \Phi) - \mu_2^2
\left[ {\rm Tr} (\widetilde{\Phi} \Phi^{\dag}) + {\rm Tr} (\widetilde{\Phi}^{\dag} \Phi) \right]
- \mu_3^2 \:  {\rm Tr} (\Delta_R
\Delta_R^{\dag})
\nonumber \\
&&
\hspace{-0.5cm}
+ \lambda_1 \left[ {\rm Tr} (\Phi^{\dag} \Phi) \right]^2 + \lambda_2 \left\{ \left[
{\rm Tr} (\widetilde{\Phi} \Phi^{\dag}) \right]^2 + \left[ {\rm Tr}
(\widetilde{\Phi}^{\dag} \Phi) \right]^2 \right\}
\nonumber \\
&&
\hspace{-0.5cm}
+ \lambda_3 \: {\rm Tr} (\widetilde{\Phi} \Phi^{\dag}) {\rm Tr} (\widetilde{\Phi}^{\dag} \Phi) +
\lambda_4 \: {\rm Tr} (\Phi^{\dag} \Phi) \left[ {\rm Tr} (\widetilde{\Phi} \Phi^{\dag}) + {\rm Tr}
(\widetilde{\Phi}^{\dag} \Phi) \right]
\nonumber \\
&&
\hspace{-0.5cm}
+ \rho_1  \left[ {\rm
Tr} (\Delta_R \Delta_R^{\dag}) \right]^2 
+ \rho_2 \: {\rm Tr} (\Delta_R
\Delta_R) {\rm Tr} (\Delta_R^{\dag} \Delta_R^{\dag})
\nonumber \\
&&
\hspace{-0.5cm}
+ \alpha_1 \: {\rm Tr} (\Phi^{\dag} \Phi) {\rm Tr} (\Delta_R \Delta_R^{\dag})
+ \left[  \alpha_2 e^{i \delta_2}  {\rm Tr} (\widetilde{\Phi}^{\dag} \Phi) {\rm Tr} (\Delta_R
\Delta_R^{\dag}) + {\rm H.c.} \right]
\nonumber \\
&&
\hspace{-0.5cm}
+ \alpha_3 \: {\rm
Tr}(\Phi^{\dag} \Phi \Delta_R \Delta_R^{\dag})
-\mu^{\prime \, 2} \, \phi_{X}^{\dagger} \, \phi_{X}+\lambda^{\prime} \left(\phi_{X}^{\dagger} \phi_{X}\right)^{2}
\nonumber \\
&&
\hspace{-0.5cm}
+\eta_{1} \left(\phi_{X}^{\dagger} \phi_{X}\right) \left[ \mbox{Tr}\left(\Delta_{L}^{\dagger} \Delta_{L}\right)\right]+\eta_2\left(\phi_{X}^{\dagger} \phi_{X}\right)
\left[\mbox{Tr}\left(\Phi^\dagger\Phi\right)\right]
+L \leftrightarrow R \; ,
\end{eqnarray}
where $\mu_{i}$ $(i=1, 2, 3)$,  $\mu^\prime$, $\lambda_{i}$ ($i=1,2,3,4$), 
   $\lambda^\prime$, $\rho_i$ $(i=1,2)$,  $\alpha_{i}$ $(i=1,2, 3)$
   and $\eta_{i}$ $(i=1,2)$ are real parameters.
The electric charge operator in our model is given by 
\begin{eqnarray}
\label{Qem}
 Q_{em}=I_{3L}+I_{3R}+\frac{Q_{BL}}{2}+\frac{Q_X}{2} ,
\end{eqnarray}
  where $I_{3L}$ ($I_{3R}$) is the diagonal generators of $SU(2)_{L}$ ($SU(2)_{R}$), 
  and $Q_{BL}$ ($Q_X$) is a $U(1)_{B-L}$ ($U(1)_X$)  charge.  
We may express the Higgs fields as 
\begin{eqnarray}
\Phi = \ \left(\begin{array}{cc}
\phi^0_1 & \phi^+_2 \\ 
\phi^-_1 & \phi^0_2\end{array}\right), \; 
\Delta_L = \left(\begin{array}{cc}
\Delta^+_L/\sqrt{2} & \Delta^{++}_L \\
\Delta^0_L & -\Delta^+_L/\sqrt{2}\end{array}\right), \; 
\Delta_R =& \left(\begin{array}{cc}
\Delta^+_R/\sqrt{2} & \Delta^{++}_R \\
\Delta^0_R & -\Delta^+_R/\sqrt{2}\end{array}
\right).
\label{eq:scalar}
\end{eqnarray}
The gauge symmetry ${\cal G}_{\rm LRX}$ is broken down to $SU(3)_c \times U(1)_{em}$ 
  by the following vacuum expectation values (VEVs): 
\begin{eqnarray}
\langle \Phi \rangle=
\left(
\begin{array}{cc}
\frac{v_1}{\sqrt{2}} & \quad 0
\\
\\
0 & \quad \frac{v_2}{\sqrt{2}} \\
\end{array}
\right) , \;
\langle \Delta_{L} \rangle=
\left(
\begin{array}{cc}
0 & \quad 0
\\
\\
\frac{v_{L}}{\sqrt{2}} & \quad 0 \\
\end{array}
\right) , \;
\langle \Delta_{R} \rangle=
\left(
\begin{array}{cc}
0 & \quad 0
\\
\\
\frac{v_{R}}{\sqrt{2}} & \quad 0 \\
\end{array}
\right) , \;
\langle \phi_{X} \rangle=\frac{v_{X}}{\sqrt{2}}  .
\end{eqnarray}
For simplicity, we choose the hierarchy among VEVs 
  such that $v_{L} \ll \sqrt{v_{1}^{2}+v_{2}^{2}} \ll  v_{X} \ll v_{R}$, 
  with $v_1=v \sin \beta$, $v_2=v \cos \beta$ and $v=246$ GeV. 
The sequence of the gauge symmetry breaking is as follows:
First, the $SU(2)_R \times U(1)_{B-L}$ symmetry is broken by $v_R$, 
  yielding large masses for $W_{R}$ and $Z_{R}$.
Next, the $U(1)_X$ symmetry is broken by $v_{X}$ and the mass of $U(1)_X$ gauge boson is generated.
The electroweak symmetry breaking down to $U(1)_{em}$ is completed by $v_1$ and $v_2$.  
In the next section, we show the mass eigenvalues and corresponding eigenstates in detail.

The Yukawa couplings of the model are given by 
\begin{eqnarray}\label{LYMJ}
-{\cal L}_Y & = &
   h_{ij}^{(\ell)} \, \overline{\psi_L^i} \, \Phi \, \psi_{R}^j
+  \widetilde{h}_{ij}^{(\ell)} \, \overline{\psi_L^i} \, \widetilde{\Phi} \, \psi_{R}^j
+  h_{ij}^{(Q)} \, \overline{Q_L^i} \, \Phi \, Q_{R}^j
+ \, \widetilde{h}_{ij}^{(Q)} \, \overline{Q_L^i} \, \widetilde{\Phi} \, Q_{R}^j
\nonumber \\
&+& \frac{1}{2} f^{\psi}_{ij} \, \overline{{\psi_L^i}^c} \, \widetilde{\Delta}_{L} \, \psi_{L}^j
    + \frac{1}{2} f^{\psi}_{ij} \, \overline{{\psi_R^i}^c} \, \widetilde{\Delta}_{R} \, \psi_{R}^j
%
+\frac{1}{2} Y^\zeta \, \phi_{X} \, \overline{{\zeta_L}^c} \, \zeta_{L}
+\frac{1}{2} Y^\zeta \, \phi_{X} \, \overline{{\zeta_R}^c} \, \zeta_{R}
+\mbox{H.c.} ,
\end{eqnarray}
where $\widetilde{\Phi}=\sigma_{2} \, \Phi^{\ast} \, \sigma_{2}$, and
   $\widetilde{\Delta}_{L(R)}=i \, \sigma_{2} \, \Delta_{L(R)}$. 
Since we impose the parity symmetry,  $\psi_L^i  \leftrightarrow \psi_R^i$, 
   $Q_L^i \leftrightarrow Q_R^i$, 
   $\Delta_{L} \leftrightarrow \Delta_{R}$, 
   $\Phi \leftrightarrow \Phi^{\dagger}$ and $\zeta_{L} \leftrightarrow \zeta_{R}$, 
   the Yukawa matrices, $h_{ij}^{(\ell)}$, $\widetilde{h}_{ij}^{(\ell)}$, 
   $h_{ij}^{(Q)}$ and $\widetilde{h}_{ij}^{(Q)}$, are Hermitian matrices. 
The Dirac mass matrices for leptons and quarks are generated by $\langle \Phi \rangle \neq 0$   
   while Majorana mass matrices for left and right-handed neutrinos are 
   generated by $\langle \Delta_{L, R}\rangle \neq 0$, respectively.    
A common Majorana mass of $m=Y^\zeta v_X/\sqrt{2}$ for $\zeta_{L,R}$ 
   is generated by $ \langle \phi_X \rangle=v_X/\sqrt{2}$. 
Along with a gauge invariant Dirac mass term for $\zeta_{L,R}$ ($M$), 
  the mass terms for $\zeta$ is given by 
\begin{eqnarray}\label{LpsichiM}
{\cal L}_{mass}
 = -
\frac{1}{2}
\left(
\begin{array}{cc}
\overline{{\zeta_L}^c} & \overline{\zeta_R}\\
\end{array}
\right)
\left(
\begin{array}{cc}
m &  M\\
M &  m \\
\end{array}
\right)
\left(
\begin{array}{c}
\zeta_L \\
{\zeta_R}^c\\
\end{array}
\right) +\mbox{H.c.} ,
\end{eqnarray}
where we set $M > m >0$. 
The mass eigenvalues are given by $M_{\pm} =M \pm m$ and 
  corresponding eigenstates, $\zeta_{\ell}$ and $\zeta_h$, are defined as 
  $P_L \zeta_{\ell}= \frac{1}{\sqrt{2}} \left( \zeta_L - {\zeta_R}^c \right)$ and 
  $P_L \zeta_{h}= \frac{1}{\sqrt{2}} \left( \zeta_L + {\zeta_R}^c \right)$, 
  $P_L$ is the left-hand projection operator. 
Thanks to the $U(1)_X$ symmetry, the lighter mass eigenstate $\zeta_\ell$ 
  is stable and identified with the Majorana fermion DM. 
In the following, we call the mass of DM $\zeta_\ell$ as $m_{DM} \equiv M_-=M-m$.

\section{Mass spectrum and eigenstates of the gauge bosons}
\label{sec:3}

Through the gauge symmetry breaking by Higgs VEVs, 
   the charged gauge bosons $W_L$ and $W_R$ in the LRSM 
   acquire their masses as 
\begin{eqnarray}
\label{massesWW_{R}}
M_{W_L} \simeq \frac{1}{2} g \, v 
\hspace{0.5cm} \mbox{and} \hspace{0.5cm}
M_{W_{R}} \simeq \frac{1}{\sqrt{2}} g \, v_R. 
\end{eqnarray}
Here, we have used the hierarchy $v \ll v_R$. 
The mass eigenstate $W_L$ is identified with the SM $W$ boson. 

Since we have four neutral gauge bosons, $A_{3L\mu}$, $A_{3R\mu}$, $B_{\mu}$ and $C_{\mu}$, 
  and they mix with each other after the symmetry breaking, 
  the analysis for their mass spectrum and eigenstates is complicated. 
According to the hierarchy, $v_L \ll v \ll v_X \ll v_R$, 
  we focus on the neutral gauge boson mass terms generated by 
  the symmetry breaking of $SU(2)_R \times U(1)_{B-L} \times U(1)_X \to U(1)_Y$: 
\begin{eqnarray}\label{Lmass}
{\cal L}_{mass}=\frac{1}{2} \, \eta^{\mu\nu}\, \left(V_{\mu}\right)^{T} \, M_{sq} \, V_{\nu} \; ,
\end{eqnarray}
where $V_{\mu}=\left( \; A_{3R\mu} \; \; B_{\mu} \; \; C_{\mu} \; \right)^T$, 
  and the mass-squared matrix is given by 
\begin{equation}\label{MatrixMassa}
M_{sq}=
\left(
\begin{array}{ccc}
g^{2} \, v_{R}^{2} & -g\, g_{BL} \, v_{R}^2 & 0
\\
-g \, g_{BL} \, v_{R}^2 &  ~g_{BL}^{2} \,v_{R}^{2}+a^2 g_{BL}^{2} \, v_{X}^2~ & -a^2 \, g_{BL} \, g_X \, v_{X}^2
\\
0 & -a^2 \, g_{BL} g_X \, v_{X}^{2} & a^2 \, g_X^2 \, v_{X}^2 \\
\end{array}
\right) .
\end{equation}
We now diagonalize the mass matrix $M_{sq}$ by a $3 \times 3$ orthogonal matrix ${\cal R}$ such that 
\begin{equation}\label{Lmassmatrixvtilde}
{\cal L}_{mass}=
 \frac{1}{2} \, \eta^{\mu\nu}\, \left(\tilde{V}_{\mu}\right)^{T} \, D_{sq} \, \tilde{V}_{\nu} ,
\end{equation}
where the mass eigenstates are defied as 
  $\tilde{V}_{\mu}=\left( \; Y_\mu \; \; X_{\mu} \; \; Z_{R{\mu}} \; \right)^T = {\cal R}^T V_\mu$, 
  and $D_{sq}={\rm diag}(\, 0, \, M_X^2, \,M_{Z_R}^2 \,)$ is the mass eigenvalue matrix with 
\begin{eqnarray}\label{massesZZ'}
M_{X} \simeq |a| \, v_{X} \, \sqrt{g_X^2 + \frac{g^2 \, g_{BL}^{2} }{g^2+g_{BL}^{2}}} 
\hspace{0.3cm} \mbox{and} \hspace{0.3cm}
M_{Z_{R}} \simeq v_{R} \, \sqrt{ g^2+g_{BL}^2}.
\end{eqnarray}
Here, we have used $v_X \ll v_R$.\footnote{
Our results of the gauge boson mass spectrum and their couplings with the fermions 
  remain the same as long as $g_{BL} v_R^2 \gg a^2 g_{BL} v_X^2, a^2 g_X v_X^2$, as expected from the form of $M_{sq}$.
}
The massless state is identified with the SM hyper-charge gauge boson $Y_\mu$, 
   while $Z_{R \mu}$ is the heavy neutral boson in the LRSM. 
To determine the couplings of the gauge boson mass eigenstates with the SM fermions and the Majorana fermion DM, 
   we need to find the form of ${\cal R}$. 
Since we set $\epsilon \equiv (v_X/v_R)^2 \ll 1$, for this purpose it is sufficient to give the form of the orthogonal matrix 
   up to ${\cal O}(\epsilon)$: 
\begin{eqnarray}\label{R}
{\cal R}=\left(
\begin{array}{ccc}
\frac{g_{BL} \, g_{X}}{\sqrt{ g^2 \, g_{BL}^{2} + g^2 \, g_X^2 + g_{BL}^2 \, g_{X}^{2}}} 
&
- \frac{g \, (g_{BL}^2+g_X^2)}{\sqrt{g_{BL}^2+g_X^2}\sqrt{ g^2 \, g_{BL}^{2} + g^2 \, g_{X}^{2}+g_{BL}^2 \, g_{X}^{2}}}
& 
0 
\\
\frac{g \, g_{X}}{\sqrt{ g^2 \, g_{BL}^{2} + g^2 \, g_{X}^{2}+g_{BL}^2 \, g_{X}^{2}}} 
&
 \frac{g_{BL} \, g_{X}^2}{\sqrt{g_{BL}^2+g_X^2}\sqrt{ g^2 \, g_{BL}^{2} + g^2 \, g_{X}^{2}+g_{BL}^2 \, g_{X}^{2}}}
& 
-\frac{g_{BL}}{\sqrt{g_{BL}^2+g_{X}^2}}
\\
\frac{g \, g_{BL}}{\sqrt{ g^2 \, g_{BL}^{2} + g^2 \, g_{X}^{2}+g_{BL}^2 \, g_{X}^{2}}} 
&
\frac{g_{BL}^2 \, g_{X}}{\sqrt{g_{BL}^2+g_X^2}\sqrt{ g^2 \, g_{BL}^{2} + g^2 \, g_{X}^{2}+g_{BL}^2 \, g_{X}^{2}}}
& 
\frac{g_{X}}{\sqrt{g_{BL}^2+g_{X}^2}}
\end{array}
\right) + {\cal O}(\epsilon) .
\end{eqnarray}
By using ${\cal R}$, we rewrite the original gauge interactions in terms of the mass eigenstates. 
It is easy to check that the hyper-charge of a field ($Q_Y$) is given by 
\begin{eqnarray}
 Q_Y=I_{3R}+\frac{Q_{BL}}{2}+\frac{Q_X}{2} , 
\end{eqnarray}
and the SM $U(1)_Y$ gauge coupling ($g_Y$) is related to $g_R=g$, $g_{BL}$ and $g_X$ by 
\begin{eqnarray}
   \frac{1}{g_Y^2}=\frac{1}{g^2} + \frac{1}{g_{BL}^2} + \frac{1}{g_X^2}.
\end{eqnarray}
Using the values of $g$ and $g_Y$ at the weak scale which are fixed by 
   $ \frac{g^2 \, g_Y^2}{g^2+g_Y^2}=e^2 \simeq \frac{4 \pi}{128}$ and 
   $\frac{g_Y^2}{g^2+g_Y^2} = \sin^2 \theta_W \simeq 0.23$, 
   we find 
\begin{eqnarray}\label{gxgbl}
g_X = \frac{0.428 \, g_{BL}}{\sqrt{ (g_{BL})^2-(0.428)^2}} ,
\end{eqnarray}
and hence $g_{BL} > 0.428$ for $g_X < \infty$. 
In the next section, we consider the perturbativity condition for $g_{BL}$ and $g_X$ up to the Planck scale
   and find more severe constraints on $g_{BL}$ and $g_X$.

In the following sections, we will investigate the DM physics. 
Since the DM particle communicates with the SM particle through the $X$-portal interaction, 
   we present the explicit forms for the couplings of the $X$-boson with the SM fermions and 
   the Majorana fermion DM. 
Using the original gauge couplings and the orthogonal matrix ${\cal R}$, 
   we find the interaction terms of the form, 
\begin{eqnarray}
  {\cal L}_{int} = -\left(
  Q_Y^{f_L} \, g_f \, \overline{f_L} \gamma^{\mu} f_L 
  + Q_Y^{f_R} \, g_f \, \overline{f_R} \gamma^{\mu} f_R
  + g_{\zeta} \, \overline{\zeta_\ell} \gamma^\mu  \gamma_5 \zeta_\ell
  \right) X_{\mu}, 
  \label{X_int}
\end{eqnarray}
where $f_L$ and $f_R$ denote the left-handed and right-handed SM fermions, respectively, 
  listed in Table~\ref{Table:1}, 
  $Q_Y^{f_{L,R}}$ are their hyper-charges, 
  we have used the Dirac fermion expression for the Majorana DM $\zeta_\ell$, and 
\begin{eqnarray}
g_{f} &=&  \frac{g \, g_{Y} }{ 0.428 } \, \frac{\sqrt{(g_{BL})^2-(0.428)^2}}{\sqrt{g^2+(g_{BL})^2}}, 
\nonumber\\ 
g_{\zeta} &=& \frac{a}{4} \, \frac{g}{g_{Y}} \, \frac{ 0.428 \, g_{BL} }{\sqrt{ (g_{BL})^2-(0.428)^2}} \, \frac{g_{BL} }{\sqrt{g^2+ (g_{BL})^2}}.
\label{gfgzeta}
\end{eqnarray}
The couplings, $g_f$ and $g_\zeta$, are determined as a function of $g_{BL}$. 
In the following analysis, we see  that our results remain the same for $a \to -a$, 
  and hence we only consider $a > 0$ without loss of generality.

\section{The perturbativitiy condition on the gauge couplings}
\label{sec:4}

We have derived the relation between $g_{BL}$ and $g_X$ in Eq.~(\ref{gxgbl}) 
  to reproduce the SM $U(1)_Y$ gauge coupling constant. 
To justify our analysis in the perturbative expansion of the model, 
  we impose a theoretical consistency condition, 
  namely, the perturbativity condition on the gauge couplings.
Let us define the condition as 
\begin{eqnarray} 
     g_{BL}(M_P) \leq 4 \pi, 
      \hspace{0.6cm} \mbox{and} \hspace{0.6cm}
     g_{X}(M_P) \leq 4 \pi,  
\end{eqnarray}
for the running gauge couplings at the reduced Planck mass, $M_P=2.43 \times 10^{18}$ GeV.

To evaluate the gauge coupling values at low energies, $\mu < M_P$, 
  we employ the renormalization group (RG) equations at the one-loop level: 
\begin{eqnarray} 
     \mu \frac{d g_{BL}}{d \mu} = \beta_{BL}(g_{BL}),  
     \hspace{0.6cm} \mbox{and} \hspace{0.6cm}
    \mu \frac{d g_{X}}{d \mu} = \beta_X(g_X).  
\end{eqnarray}
With the particle content in Table~\ref{Table:1}, the beta functions of $\beta_{BL}$ and $\beta_X$ 
  are  calculated to be 
\begin{eqnarray}\label{beta}
\beta_{BL}= \left(\frac{28+a^2}{6} \right) \, \frac{g_{BL}^{3}}{16\pi^2}, 
\hspace{0.6cm} \mbox{and} \hspace{0.6cm}
\beta_{X}= \left( \frac{a^2}{6} \right) \, \frac{g_{X}^{3}}{16\pi^2} .
\end{eqnarray}
Solving the RG equations, we find the maximum values of $g_{BL}$ and $g_X$ at $v_R$,  
\begin{eqnarray}
\left. g_{BL} \right|_{max} = \frac{4\pi}{ \sqrt{1+\left(\frac{28+a^2}{3}\right) \ln \left[\frac{M_{P}}{v_R} \right] } }
\hspace{0.5cm} \mbox{and} \hspace{0.5cm}
\left. g_{X} \right|_{max} = \frac{4\pi}{ \sqrt{ 1+\frac{a^2}{3} \ln \left[\frac{M_{P}}{v_R} \right] } } \; .
\end{eqnarray}
In this paper, we set $v_R=10^5$ GeV. 
Since the $v_R$ value is not far from the electroweak scale, we approximate $g_Y(v_R)=g_Y(v)$. 
Note that the relation between $g_{BL}$ and $g_X$ of Eq.~(\ref{gxgbl}) indicates that
   the maximum value of $\left. g_{BL}\right|_{max} $ ($\left. g_X\right|_{max} $) 
   corresponds to the minimum value of $\left.g_X\right|_{min} $ ($\left. g_{BL}\right|_{min} $). 
Similarly, from Eq.~(\ref{gfgzeta}), 
     $\left. g_f\right|_{max}$ and  $\left. g_\zeta\right|_{min}$ 
   ($\left. g_f\right|_{min}$ and  $\left. g_\zeta\right|_{max}$)
    correspond to $\left. g_{BL}\right|_{max}$  ($\left. g_{BL}\right|_{min} $).

\begin{figure}[t]
\centering
\includegraphics[width=0.47\textwidth]{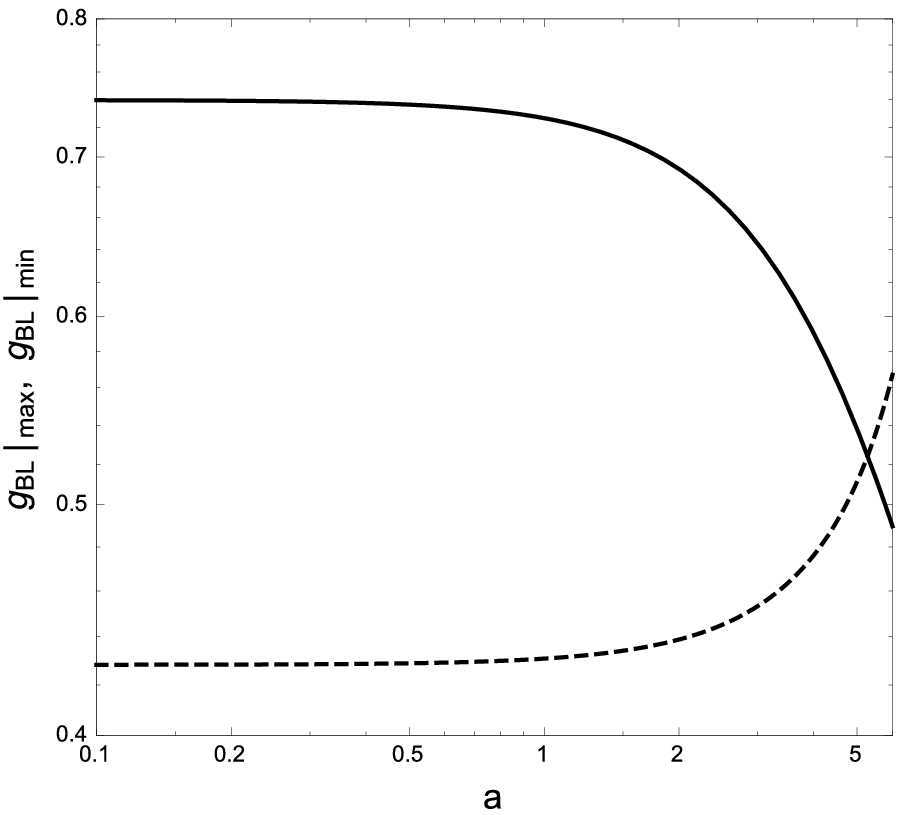}\; \; \;
\includegraphics[width=0.47\textwidth]{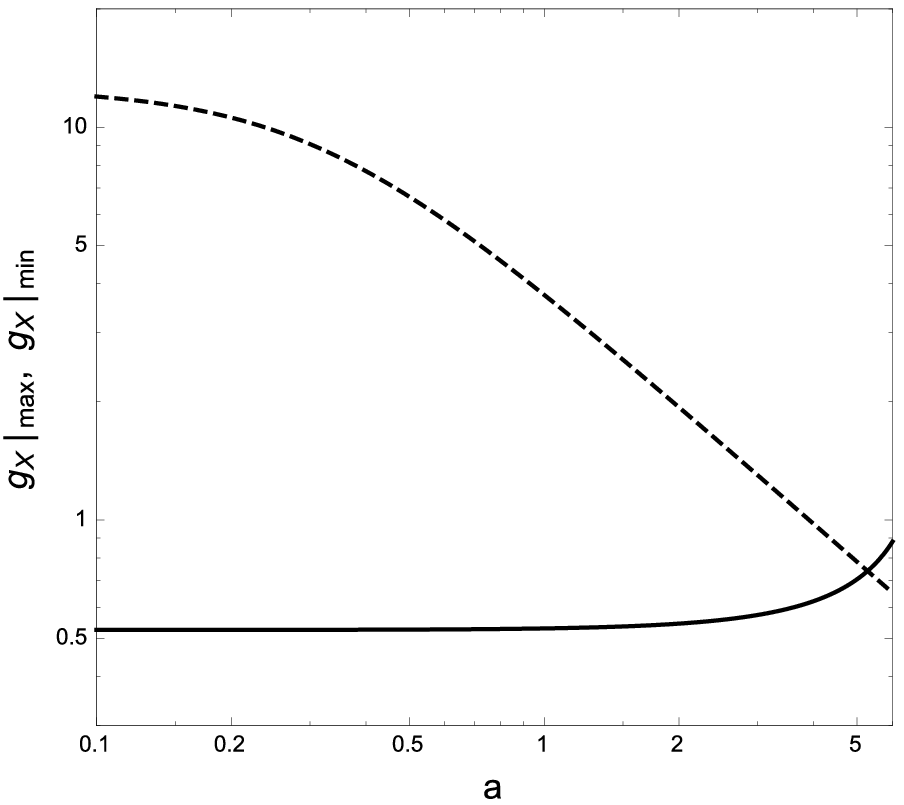}
\caption{
{\it Left Panel}: The solid and dashed lines depict the maximum
  and minimum values of $g_{BL}$ at $\mu=v_R=10^5$ GeV, respectively, 
  as a function of $a$.  
{\it Right Panel}: The minimum and maximum values of $g_{X}$ as a function of $a$, 
  which correspond to the maximum and minimum values of $g_{BL}$ 
  shown in Left Panel. 
}
\label{gblgx}
\end{figure}
\begin{figure}[t]
\centering
\includegraphics[width=0.47\textwidth]{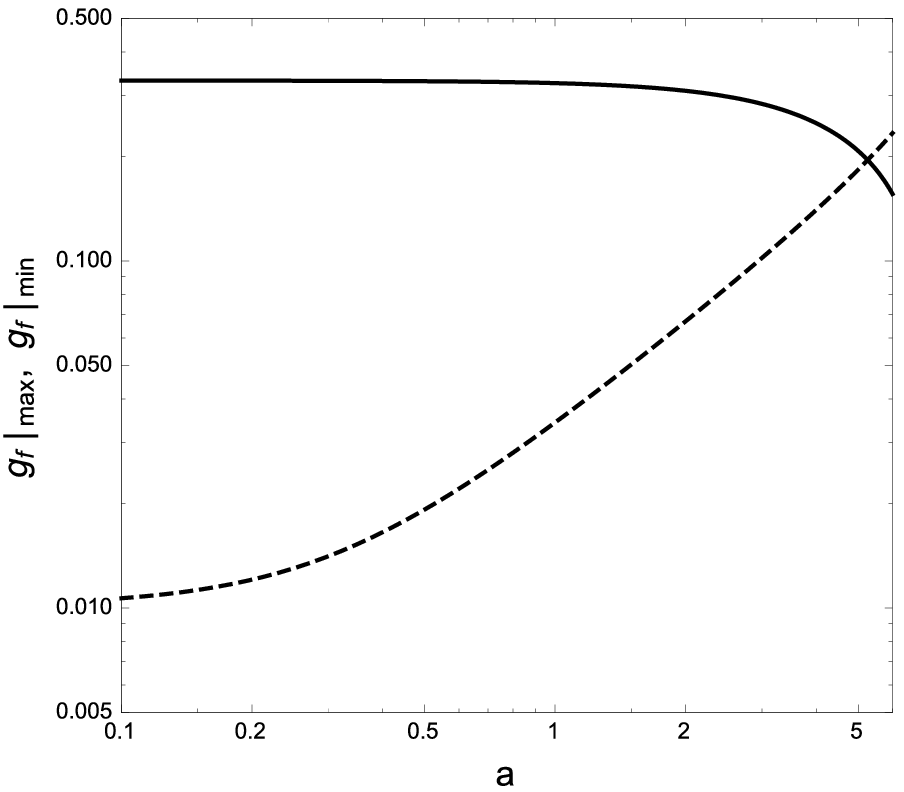} \; \; \;
\includegraphics[width=0.47\textwidth]{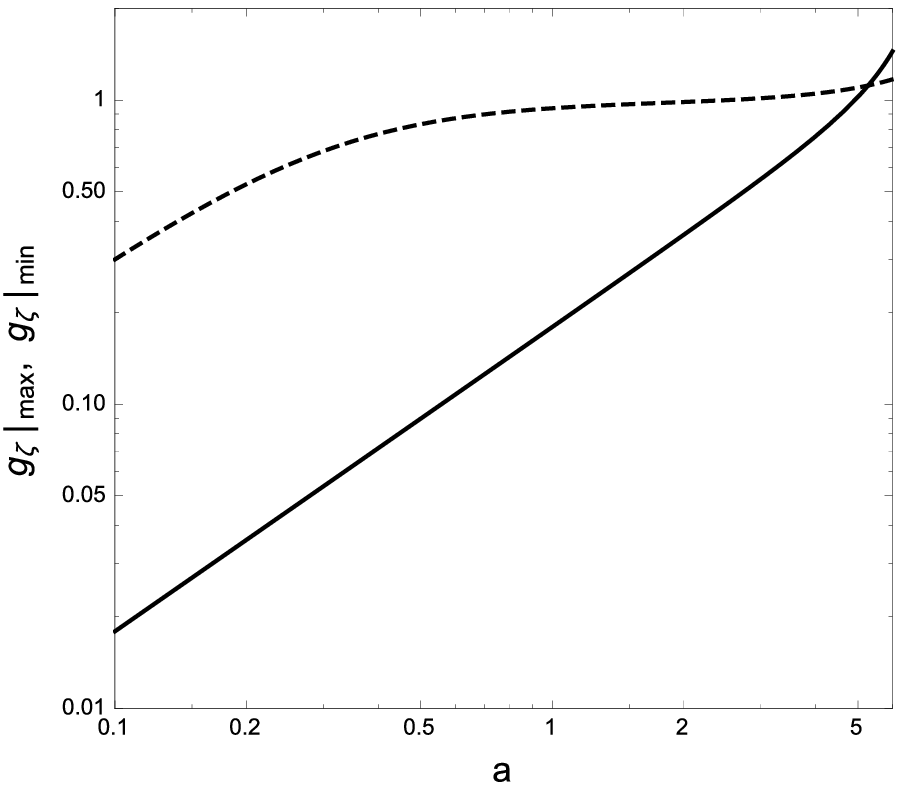}
\caption{
{\it Left Panel}:
The maximum (solid line) and minimum (dashed line) values of $g_f$ as a function of $a$, 
   corresponding to $\left.g_{BL}\right|_{max}$ and  $\left.g_{BL}\right|_{mix}$.
{\it Right Panel}:
The maximum (dashed line) and minimum (solid line) values of $g_\zeta$ as a function of $a$,
   corresponding to $\left.g_{BL}\right|_{max}$ and  $\left.g_{BL}\right|_{mix}$.
}
\label{Fig:gfgzeta}
\end{figure}

\begin{table}[t]
\centering
\begin{tabular}{|c||c|c|c|c|}
\hline
  $a$   & \quad $g_{BL}|_{min}$ \quad & \quad $g_{X}|_{max}$ \quad & \quad $g_{f}|_{min}$ \quad & \quad $g_{\zeta}|_{max}$ \quad \\
\hline
\hline
 $0.3$   &$0.428$ & $9.06$ & $0.0141$ & $0.680$  \\
\hline
 $1$   & $0.431$ & $3.74$ & $0.0342$ & $0.941$  \\
\hline
 $2$   & $0.439$ & $1.94$ & $0.0670$ & $0.986$  \\
\hline
 $3$   & $0.453$ & $1.30$ & $0.102$ & $1.01$  \\
\hline
 $5$   & $0.511$ & $0.783$ & $0.183$ & $1.10$  \\
\hline
\end{tabular} 
\begin{tabular}{|c||c|c|c|c|}
\hline
  $a$   & \quad $g_{BL}|_{max}$ \quad & \quad $g_{X}|_{min}$ \quad & \quad $g_{f}|_{max}$ \quad & \quad $g_{\zeta}|_{min}$ \quad \\
\hline
\hline
 $0.3$   &$0.738$ & $0.525$ & $0.331$ & $0.0538$  \\
\hline
 $1$   & $0.727$ & $0.530$ & $0.326$ & $0.179$  \\
\hline
 $2$   & $0.692$ & $0.545$ & $0.309$ & $0.361$  \\
\hline
 $3$   & $0.644$ & $0.573$ & $0.284$ & $0.550$  \\
\hline
$5$   & $0.538$ & $0.706$ & $0.208$ & $1.02$  \\
\hline
\hline
\end{tabular}
\caption{
The minimum and maximum values of $g_{BL}$, $g_X$, $g_{f}$ and $g_{\zeta}$ 
  for various values of $a$. 
}
\label{Table:2}
\end{table}

In Fig.~\ref{gblgx}, we show $\left. g_{BL}\right|_{max, min}$ (left panel) 
     and $\left. g_X\right|_{max, min}$ (right panel) as a function of $a$.  
The value of $a$ is restricted to be $0 < a \leq 5.28$ from the consistency, 
   $\left. g_{BL}\right|_{min} \leq \left. g_{BL}\right|_{max}$. 
In Fig.~\ref{Fig:gfgzeta}, we plot $\left. g_f \right|_{max, min}$ (left panel) 
   and $\left. g_\zeta \right|_{min, max}$ (right panel) 
   as a function of $a$, corresponding to $\left. g_{BL}\right|_{max, min}$  
   in the left panel of Fig.~\ref{gblgx}. 
For several $a$ values, we list the maximum and minimum values of $g_{BL}$, $g_X$, $g_f$ and $g_\zeta$
   in Table~\ref{Table:2}.

\section{LHC constraints}
\label{sec:5}

In the gauge extension of the Standard Model, a new gauge boson appears. 
If kinematically allowed, such a gauge boson can be produced at many experiments, 
  in particular, high energy collider experiments like the LHC experiment.  
The ATLAS and the CMS collaborations have been searching for a narrow resonance 
  with a variety of final states, among which the results with dilepton final states 
  provide the most severe constraints (unless the branching ratio of a resonance state 
  is significantly suppressed). 
The ATLAS \cite{Aad:2019fac} and the CMS \cite{CMS:2019tbu} collaborations have reported their final results with the full LHC Run-2 data, 
  which very severely constrain the production cross section of a charge-neutral vector boson (so-called $Z^\prime$ boson). 
For example, let us consider the LHC search for the sequential SM $Z^\prime$ boson ($Z^\prime_{SSM}$), 
  whose interaction is exactly the same as that of the SM $Z$ boson. 
Since no indication of $Z^\prime_{SSM}$ productions has been observed at the LHC Run-2, 
  the lower bound on $Z^\prime_{SSM}$ boson mass has been obtained 
  as $M_{Z^\prime_{SSM}} \geq 5.1$ TeV by the ATLAS results \cite{Aad:2019fac} with 139/fb integrated luminosity 
  and $M_{Z^\prime_{SSM}} \geq 5.15$ TeV by the CMS results \cite{CMS:2019tbu} with 140/fb integrated luminosity. 
Our model includes 3 new gauge bosons, namely, $W_R$, $Z_R$ and $X$. 
Since we set $v_R =10^5$ GeV $\gg v_X$, $W_R$ and $Z_R$ are too heavy to be produced at the LHC. 
In this section, we consider the production of the $X$ boson at the LHC and 
  the current constraints on the $X$ boson production by the narrow resonance search 
  with dilepton final states.

We first calculate the $X$ boson partial decay width into a pair of SM chiral fermions 
  ($f_{L,R}$) (neglecting their masses) and a pair of DM particles $\zeta_\ell$: 
\bea
 \Gamma({X\to {\overline{f_{L(R)}}} \,  f_{L(R)}}) &\ = \ & N_c \; \frac{g_f^2 }{24 \pi} (Q_Y^{f_{L(R)}})^2 \, M_X, \nonumber \\
 \Gamma(X \to \zeta_\ell \, \zeta_\ell)
 &\ = \ & \frac{g_\zeta^2 }{24\pi} M_X  \left(1-\frac{4 m_{DM}^2}{M_X^2}\right)^{3/2}, 
\label{Xwidths}
\eea 
where $N_c = 1 (3)$ is the color factor for a SM lepton (quark), 
  and we have assumed that the the $X$ boson decay into $\zeta_h$ is kinematically forbidden, for simplicity. 
The total decay width of the $X$ boson is the sum of partial widths to all SM fermions and the DM particles. 
As we will discuss in the next section, $m_{DM} \simeq M_X/2$ is required to reproduce 
  the observed DM relic density, and the contribution of  $\Gamma(X \to \zeta_\ell \, \zeta_\ell)$ 
  to the total decay width is found to be negligibly small. 
Thus, we neglect $\Gamma(X \to \zeta_\ell \, \zeta_\ell)$ in our LHC analysis.

In evaluating the $X$ boson production cross section at the LHC, 
   we first notice that the LHC Run-2 constraints are very severe on $Z^\prime$ boson productions, 
   so that we expect that the $X$ boson coupling with the SM fermions is constrained to be $g_f \ll1$. 
This means that the total $X$ boson decay width ($\Gamma_X$) is very narrow, 
   and we use the narrow width approximation in our calculation. 
In this approximation, the $X$ boson production cross section at the parton level 
  ($q \bar{q} \to X$) is given by
\bea
  \hat{\sigma}(\hat{s}) \ = \ \frac{4 \pi^2}{3} \frac{\Gamma(X \to q \bar{q})}{M_X} 
   \, \delta (\hat{s}-M_X^2) , 
 \label{Xsec_parton}  
\eea             
where $\hat{s}$ is the invariant mass squared of the colliding partons (quarks).  
With this $\hat{\sigma}$,  the cross section of the process $pp \to X$ at the LHC Run-2 with $\sqrt{s}=13$ TeV 
   is calculate by 
\bea 
   \sigma( pp \to X) \ = \ 2 \sum_{q, \, \bar{q}} \int_0^1 d x \int_0^1 dy  \, f_q(x, Q) \,  f_{\bar{q}} (y, Q) \, \hat{\sigma}(x y s), 
 \label{Xsec_X}  
\eea
  where$f_q$ ($ f_{\bar{q}}$) is the parton distribution function (PDF) for a quark (anti-quark).   
For the PDFs, we employ CTEQ6L \cite{Pumplin:2002vw} with a factorization scale $Q=M_X$, for simplicity. 
  
\begin{figure}[t]
\centering
\includegraphics[width=0.47\textwidth]{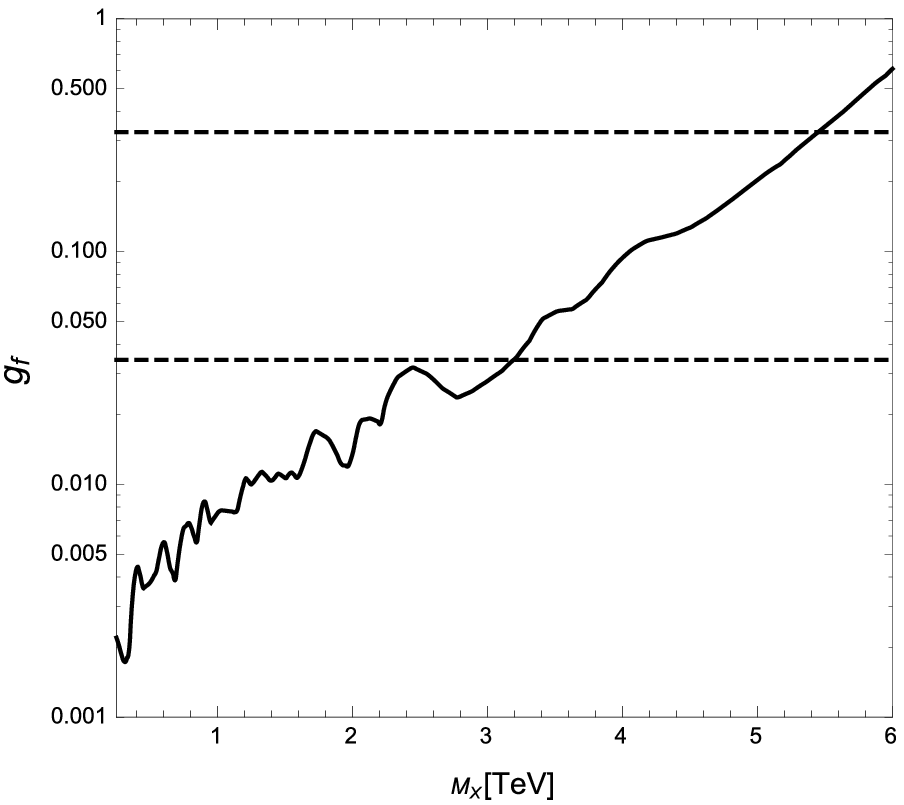} \; \; \;
\includegraphics[width=0.45\textwidth]{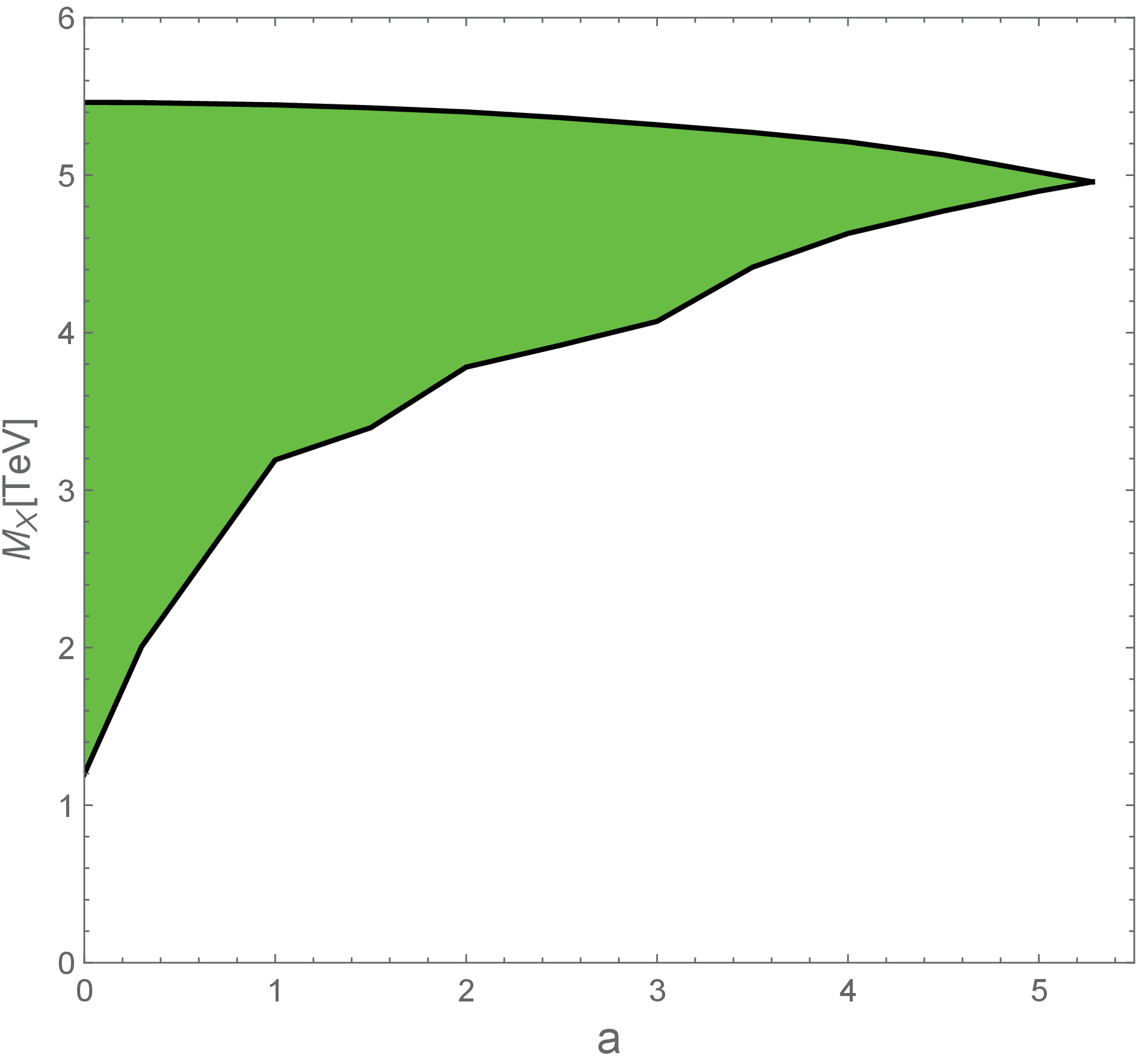}
\caption{
{\it Left Panel}:
The upper bound on $g_f$ (solid line) as a function of $M_X$ from the ATLAS results \cite{Aad:2019fac}. 
Along with the LHC bound, we also show the perturbativity condition on $g_f$.  
The two horizontal lines depict $\left. g_f\right|_{min}$ and $\left. g_f\right|_{max}$ for $a=1$. 
Combining the LHC bound and the perturbativity condition, 
   we find the allowed region, $3.19 \leq M_X[{\rm TeV}] \leq 5.45$ for $a=1$. 
{\it Right Panel}:
The allowed region of $M_X$ (green shaded) for various values of $a$ 
   after combining the LHC bound and the perturbativity condition. 
}
\label{Fig:LHC}
\end{figure}

We obtain $\sigma(pp\to X) \, {\rm BR}(X \to \ell^+\ell^-)$ as a function of $g_f$ and $M_X$. 
In the narrow decay width approximation, this cross section is proportional to $g_f^2$. 
Comparing our cross section with the upper bound by the ATLAS collaboration \cite{Aad:2019fac} 
   for fixed $M_X$ values, we obtain the upper bound on $g_f$ as a function of $M_X$. 
Our result is shown in Fig.~\ref{Fig:LHC}.   
The left panel depicts the upper bound on $g_f$ (solid line). 
We also show $g_f\left. \right|_{min}$ and $g_f\left. \right|_{max}$ for $a=1$, as an example, 
  from the perturbativity condition discussed in the previous section. 
Combining the LHC bounds and perturbativey condition, we find the allowed region, $3.19 \leq M_X[{\rm TeV}] \leq 5.45$ for $a=1$. 
For various $a$ values, we identify the allowed region of $M_X$, which is shown in the right panel 
  of Fig.~\ref{Fig:LHC} (green shaded region).

\section{Cosmological constraint}
\label{sec:6}

The DM particle $\zeta_\ell$ in our model can communicate with the SM particles
   through its interactions with the $X$ and $Z_R$ bosons and the Higgs bosons. 
For simplicity, we assume that the mixings of $\phi_X$ with $\Phi$ and $\Delta_{L,R}$ 
   are very small and hence Higgs boson mediated interactions are unimportant 
   for the DM physics. 
Since we have set $v_R=10^5$ GeV and the $Z_R$ boson is very heavy, 
   the DM particle communicates with the SM particles
   mainly through its interaction with the $X$ boson given in Eq.~(\ref{X_int}).  
In this section, we investigate this ``$X$-portal DM'' scenario 
  to identify the allowed parameter region from the cosmological constraint, 
  namely, the observed DM relic density.

In the early universe, the DM particle $\zeta_\ell$ was in thermal equilibrium 
   with the SM particles through its $X$ boson interaction. 
Due to the expansion of the universe, the DM particle decoupled form the SM particle thermal plasma
   at the freeze-out time in the early universe and then the total number of the DM particles in the universe is fixed. 
At the freeze-out time, we consider two main processes for the DM pair annihilations:  
  (i) $\zeta_\ell \, \zeta_\ell \to X \to f_{SM} \, \overline{f_{SM}}$ and 
  (ii) $\zeta_\ell \, \zeta_\ell \to X \, X$, 
  where $f_{SM}$ represents an SM fermion. 
The annihilation cross sections are controlled by four parameters: 
  $m_{DM}$, $M_X$, $g_f$ and $g_\zeta$. 
With Eq.~(\ref{gfgzeta}), we use $m_{DM}$, $M_X$, $g_{BL}(\mu=v_R)$ and $a$
  as free parameters in our DM physics analysis.  
As we have discussed in Secs.~\ref{sec:4} and \ref{sec:5}, 
  once we fix a value for $a$, the range of $g_{BL}$ is constrained by the perturbativity condition, 
  and combining it with the LHC constraints, the range of $M_X$ is also restricted. 
Note that for $m_{DM} \simeq M_X/2$, 
   the process (i) dominates the annihilation cross section through $X$ boson resonance effect. 
The process (ii)  is relevant only for $m_{DM} > M_X$.

For evaluating the DM relic density, we solve the Boltzmann equation 
(for a review, see Refs.~\cite{Kolb:1990vq, Bertone:2004pz}): 
\bea 
\frac{d Y}{d x}
= -\frac{s(m_{DM})}{H(m_{DM})} \,  \frac{\langle\sigma v_{rel} \rangle}{x^2} \, (Y^2-Y_{EQ}^2) ,
\label{Y:Boltzmann1}
\eea 
where the (photon) temperature of the universe ($T$) is normalized by $x=m_{DM}/T$, 
   $s(m_{DM})$ and $H(m_{DM})$ are the entropy density and the Hubble parameter at $T=m_{DM}$, respectively, 
   $Y$ is the yield of DM particle (the ratio of the DM number density to the entropy density), 
   $Y_{EQ}$ is the yield of the DM particle in thermal equilibrium, 
  and $\langle \sigma v_{rel} \rangle$ is the thermal average of the DM annihilation cross section ($\sigma$) 
  times relative velocity ($v_{rel}$). 
Explicit formulas of $s$, $H$ and $Y_{EQ}(x)$ are given as follows: 
\begin{eqnarray}
s(T) &=& \frac{2\pi^2}{45} g_{\ast} \, T^3 = \frac{2\pi^2}{45} \, g_{\ast} \left(\frac{m_{DM}}{x}\right)^3, 
\nonumber \\
H(T) &=& \sqrt{\frac{\pi^2}{90} \, g_{\ast}} \, \frac{T^2}{M_P}, \nonumber \\
s \, Y_{EQ} &=& \frac{g_{DM}}{2 \pi^2} \left( \frac{m_{DM}}{x} \right)^3  x^2 \, K_2(x),   
\label{functions}
\end{eqnarray}
where $g_{DM}=2$ is the number of degrees of freedom for the Majorana fermion DM $\zeta_\ell$, 
   $g_\ast$ is the effective total number of degrees of freedom for the particles in thermal equilibrium 
   (in our analysis, we use $g_\ast=106.75$ for the SM particles),  
   and $K_2$ is the modified Bessel function of the second kind.   
The thermal averaged annihilation cross section is given by 
\bea 
\langle \sigma v_{rel} \rangle = \left(s Y_{EQ} \right)^{-2} \, g_{DM}^2 \,
  \frac{m_{DM}}{64 \pi^4 x} 
  \int_{(2 m_{DM})^2}^\infty  ds \,  
  2 (s- (2 m_{DM})^2) \, \sigma(s)  \, \sqrt{s} K_1 \left(\frac{x \sqrt{s}}{m_{DM}}\right) , 
\label{ThAvgSigma}
\eea
where $\sigma(s)$ is the DM pair annihilation cross section, and $K_1$ is the modified Bessel function of the first kind. 
Solving the Boltzmann equation with the initial condition $Y(x) = Y_{EQ}(x)$ for $x \ll 1$, 
  the DM relic density at present is evaluated by
\bea
 \Omega_{\rm DM} h^2 = \frac{m_{DM} \, s_0 \, Y(x_0)}{\rho_c/h^2}, 
 \label{Eq:Omega}
\eea
where $s_0=2890 \, \mbox{cm}^3$ is the entropy density of the present universe, 
  $\rho_c/h^2=1.05 \times 10^{-5}$ GeV/cm$^3$ is the critical density, 
  and $Y(x_0)$ is the DM yield at present ($x_0 \gg 1$).   
We impose the cosmological constraint, namely, 
  $\Omega_{\rm DM} h^2 = 0.12$ to reproduce the observed DM relic density 
  set by the Planck 2018 measurements \cite{Aghanim:2018eyx}.

We first consider the parameter region $m_{DM} \simeq M_X/2$, 
  in which case the DM pair annihilation process (i) dominates the annihilation cross section 
  by the $X$ boson resonance effect. 
For the process $\zeta_\ell \, \zeta_\ell \to X \to f_{L(R)} \, \overline{f_{L(R)}}$, we find the annihilation cross section
  of the form: 
\begin{equation}\label{sigma0}
\sigma(s) = \frac{g_{\zeta}^2 \, g_f^2}{48 \pi} \, \frac{\sqrt{s \, \left(s-4m_{DM}^{2}\right)}}{\left(s-M^2_X\right)^2+M_{X}^{\, 2} \, \Gamma_{X}^{\, 2} }
 \left( \sum_{f_L}  N_c \,  (Q_Y^{f_{L}})^2 + \sum_{f_R}  N_c \,  (Q_Y^{f_{R}})^2
 \right), 
\end{equation}
where we have neglected the SM fermion masses 
  since $M_X$ is constrained to be in the range of $1 \lesssim M_X[{\rm TeV}] \lesssim 5.5$ 
  as discussed in the previous section (see the right panel of Fig.~\ref{Fig:LHC}).

\begin{figure}[t]
\centering
\includegraphics[width=0.46\textwidth]{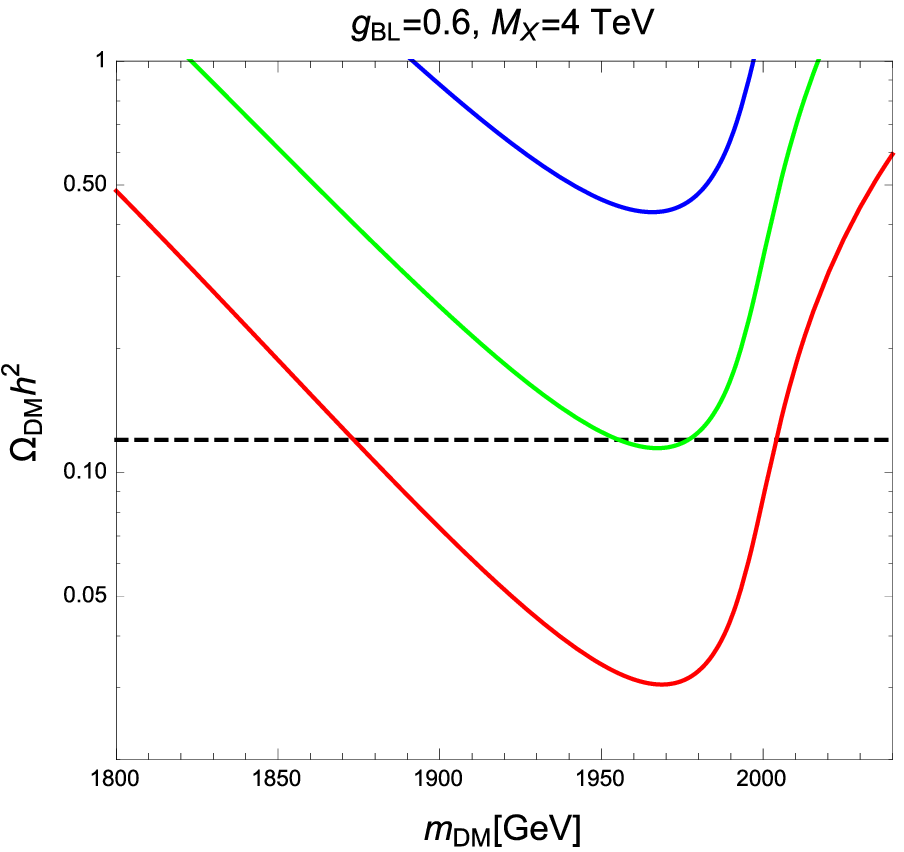} \; \; \;
\includegraphics[width=0.46\textwidth]{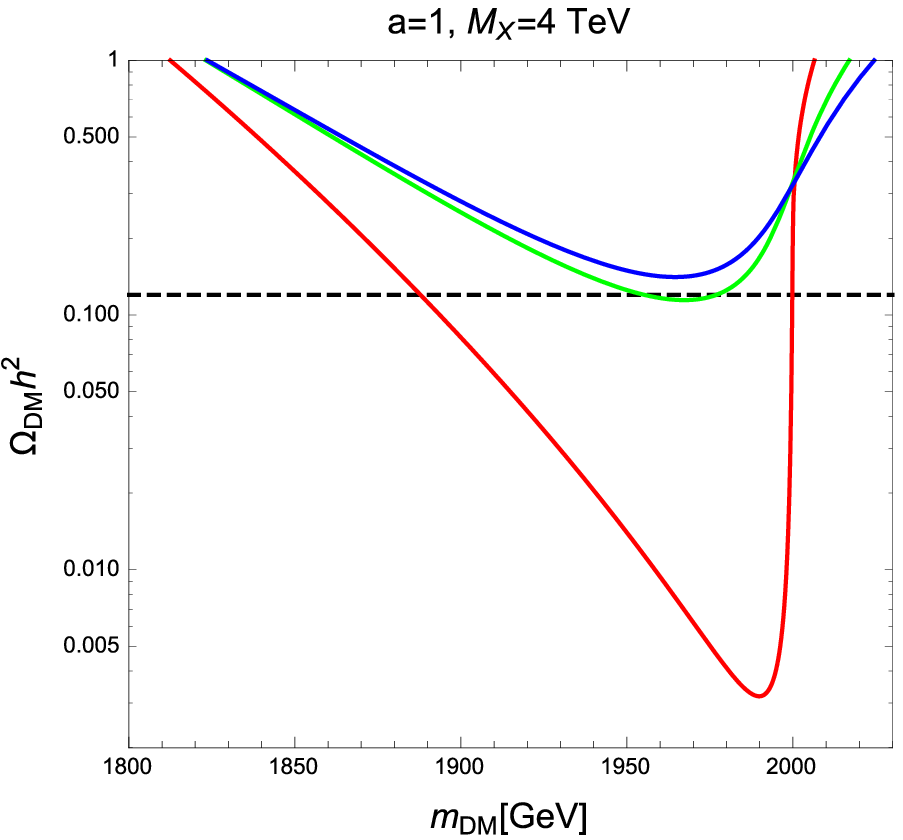}
\caption{
{\it Left Panel}:
The resultant DM relic densities for $a=0.5$ (blue), $a=1$ (green) and $a=2$ (red), 
  respectively, as a function fo $m_{DM}$, 
  along with the observed value (dashed horizontal line) of $\Omega_{DM} h^2=0.12$.  
In this analysis, we have fixed  $g_{BL}=0.6$ and $M_{X}= 4$ TeV. 
{\it Right Panel}:
The resultant DM relic densities for $a=1$ and $M_X=4$ TeV, 
  along with the observed value (dashed horizontal line) of $\Omega_{DM} h^2=0.12$.  
The blue, green and red lines from top to bottom, respectively, 
 correspond to the results with 
 $g_{BL}=\left. g_{BL}\right|_{max}=0.727$, $0.6$ and $\left. g_{BL}\right|_{min}=0.431$
 (see Table~\ref{Table:2}). 
}
\label{Fig:Omega}
\end{figure}

Using this in Eq.~(\ref{ThAvgSigma}), we numerically solve the Boltzmann equation of Eq.~(\ref{Y:Boltzmann1})
  and then evaluate the DM relic density by Eq.~(\ref{Eq:Omega}).
In Fig.~\ref{Fig:Omega}, we show the resultant DM relic densities as a function of $m_{DM}$. 
The left panel shows $\Omega_{DM} h^2$ 
  for $a=0.5$ (blue line), $a=1$ (green line) and $a=2$ (red line)
  from top to bottom, respectively, as a function of $m_{DM}$, 
  along with the observed value (dashed horizontal line) of $\Omega_{DM} h^2=0.12$.  
In this analysis, we have fixed $g_{BL}=0.6$ and $M_{X}= 4$ TeV. 
We see that the observed DM relic density can be reproduced for a suitable choice of  
  $m_{DM} \simeq M_X/2$ for $a \gtrsim 1$. 
For $a=1$ and $M_{X}= 4$ TeV, 
 we show the resultant $\Omega_{DM} h^2$ in the right panel. 
The blue, green and red lines from top to bottom, respectively, 
 correspond to the results with 
 $g_{BL}=\left.g_{BL}\right|_{max}=0.727$, $0.6$ and $\left.g_{BL}\right|_{min}=0.431$
 (see Table~\ref{Table:2}). 
Our results indicate that an enhancement of the DM annihilation cross section by 
  the $X$ boson resonance effect is crucial for reproducing the observed DM relic density. 
We have checked that for the parameters used in this analysis, 
  the annihilation cross section of the process (ii) is negligibly small 
  compared with the process (i).

\begin{figure}[t]
\centering
\includegraphics[width=0.47\textwidth]{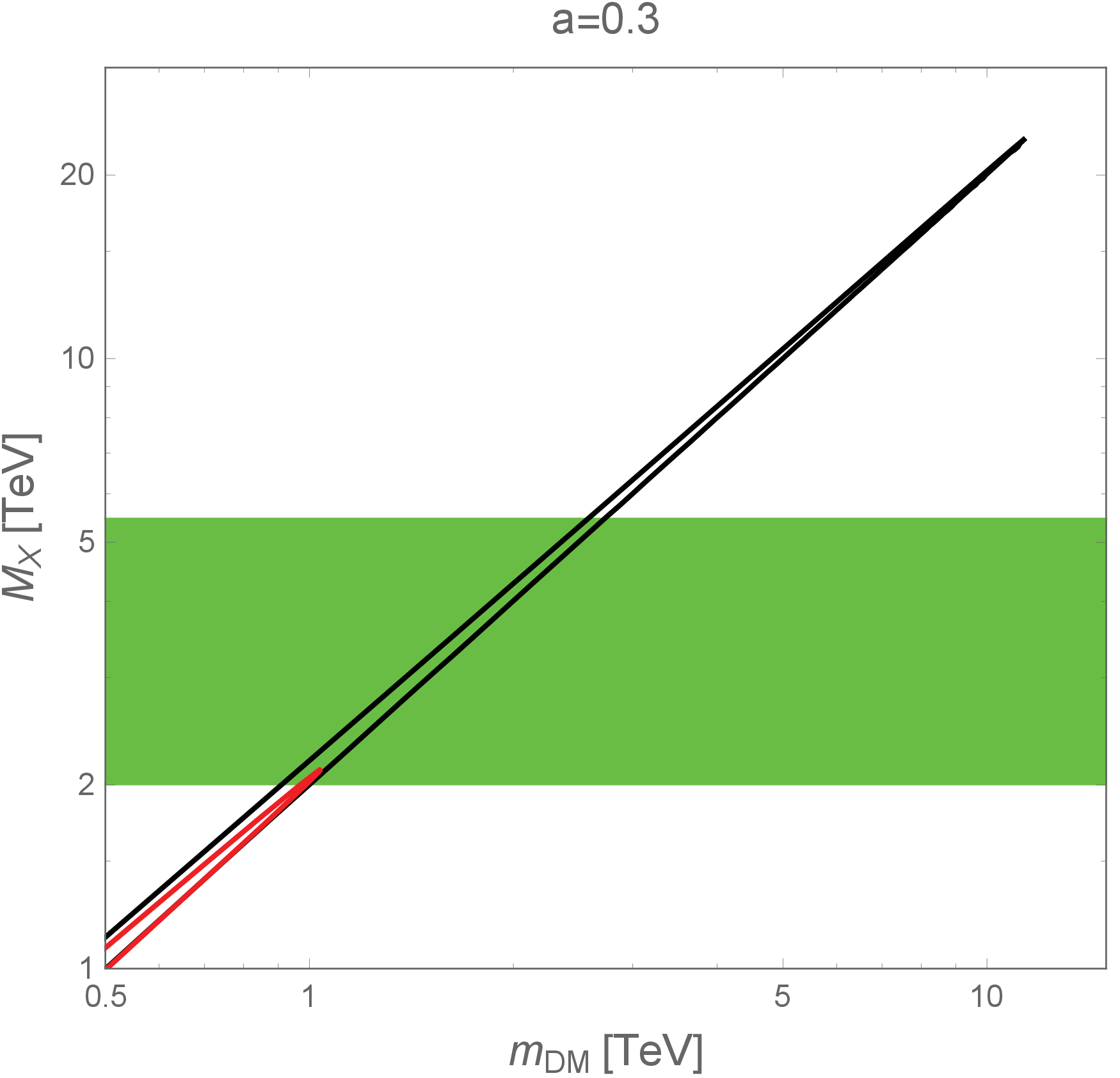} \; \; \; 
\includegraphics[width=0.47\textwidth]{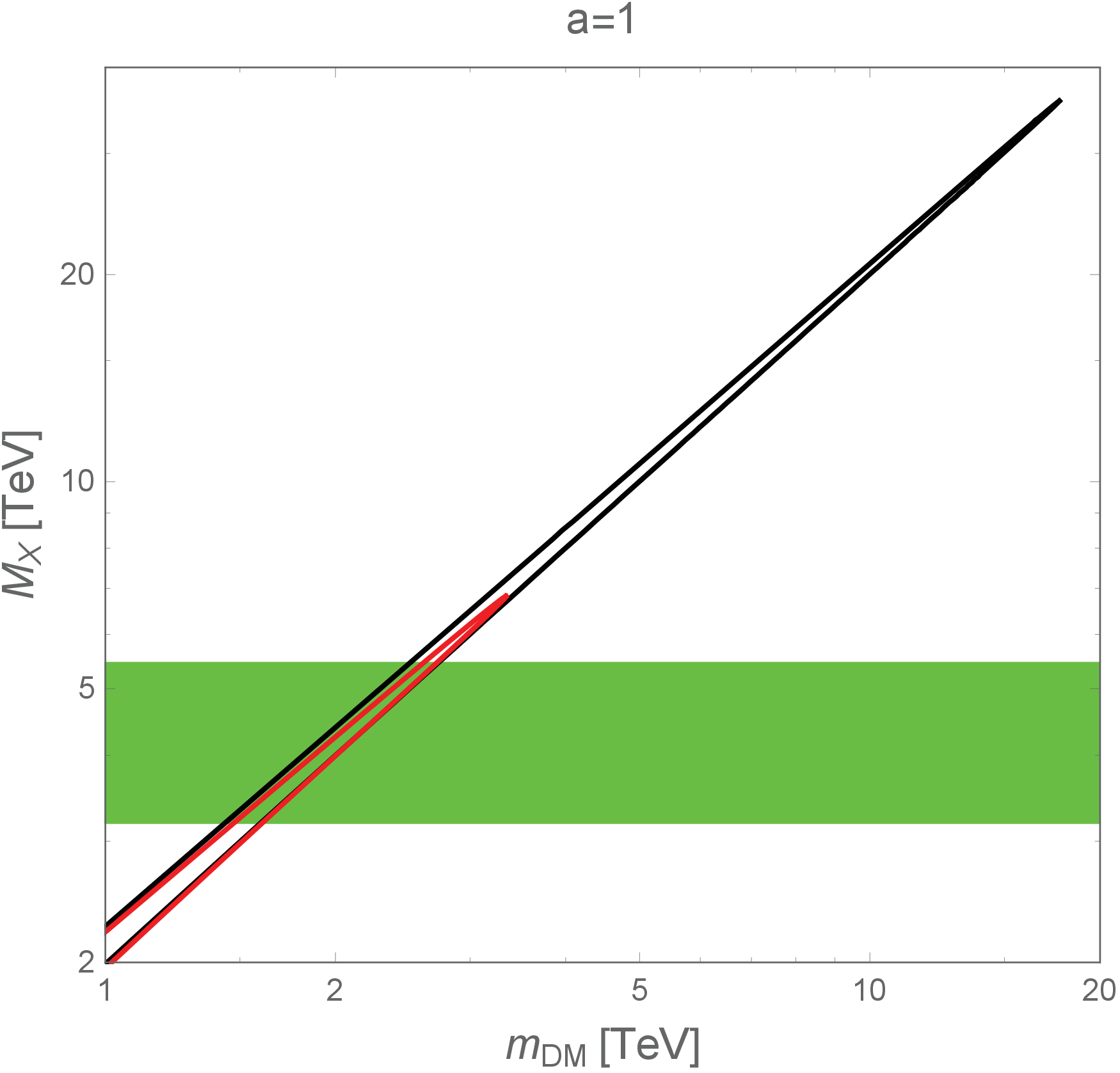} \;\; \;\\
\includegraphics[width=0.47\textwidth]{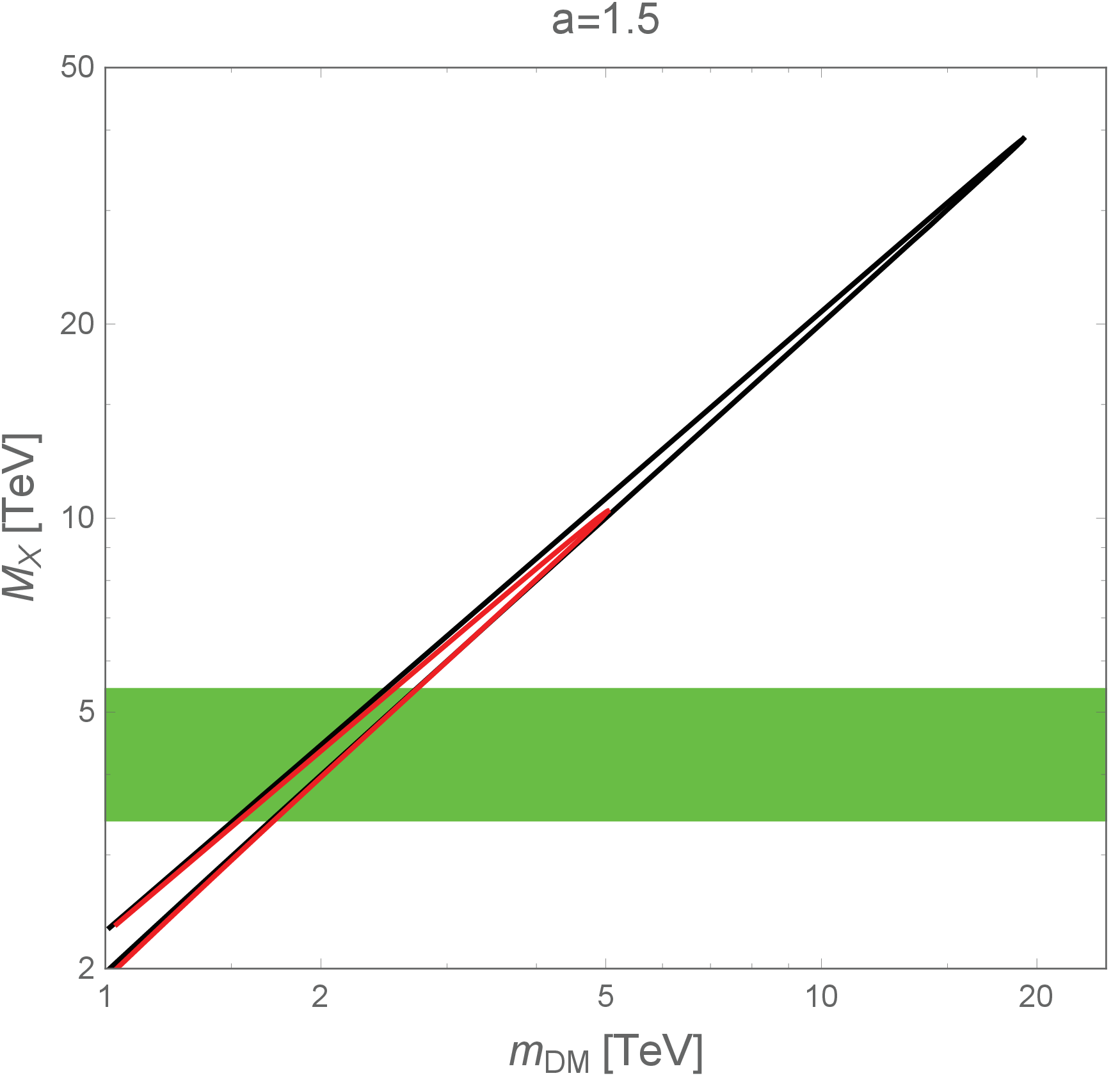} \; \; \;
\includegraphics[width=0.47\textwidth]{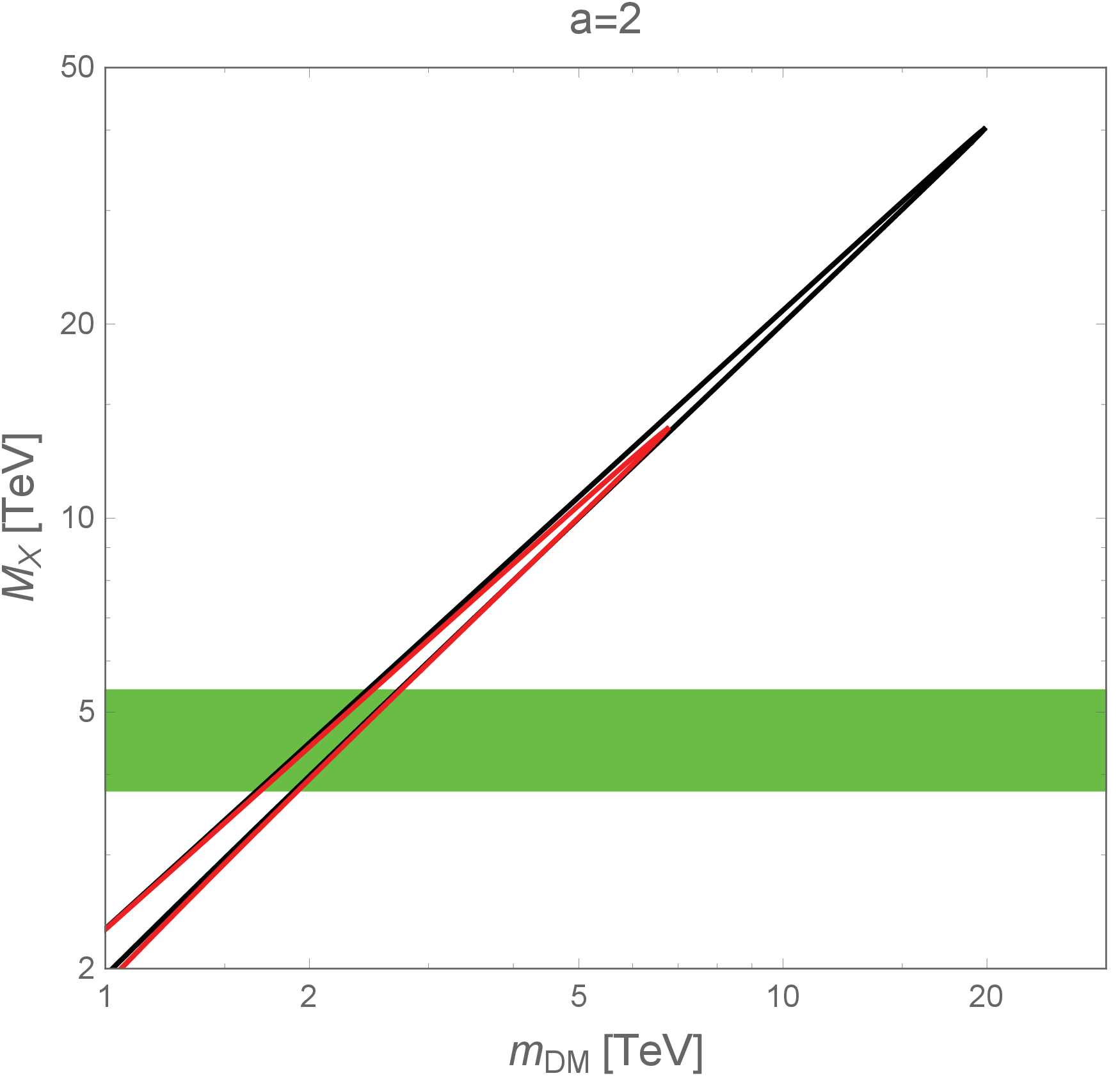} 
\caption{
The parameter region to reproduce the observed DM relic density for various $a$ values. 
The black and red solid lines correspond to the results for $\left. g_{BL}\right|_{min}$ and $\left. g_{BL}\right|_{max}$, 
   respectively, along which $\Omega_{DM} h^2=0.12$ is reproduced.  
The green shaded regions depict the ranges of $M_X$ 
  which simultaneously satisfy the perturbativity condition 
  and the LHC Run-2 constraints.
}
\label{Fig:DMregions}
\end{figure}

As can be seen from Fig.~\ref{Fig:Omega}, for fixed values of $a$, $g_{BL}$ and $M_X$, 
   the DM mass to reproduce the observed DM relic density 
   is read off from an intersection of the solid line and the dashed line. 
In Fig.~\ref{Fig:DMregions}, we show the relations between $m_{DM}$ and $M_X$
  for $a=0.3$, $1$, $1.5$ and $2$, respectively, 
  so as to reproduce the observed DM relic density. 
In each panel, the black and red solid lines correspond to the results for $\left. g_{BL}\right|_{min}$ and $\left. g_{BL}\right|_{max}$, 
  respectively, along which $\Omega_{DM} h^2=0.12$.  
The green shaded regions depict the ranges of $M_X$ 
  which simultaneously satisfy the perturbativity condition 
  and the LHC Run-2 constraints (see the right panel of Fig.~\ref{Fig:LHC}).  
Now we can see that the allowed parameter region is very limited after combining all the constraints.

Next we consider the case that the process (ii) $\zeta_\ell \, \zeta_\ell \to X \, X$ dominates 
  the annihilation cross section. 
We can see from Fig.~\ref{Fig:Omega}, the cross section of the process (i) sharply drops 
  as $m_{DM}$ goes away from the $X$ boson resonance point. 
For $m_{DM} > M_X$, the process (ii) can dominate the annihilation cross section 
  if $g_\zeta$ is sufficiently large.

Since the process (ii) is an $s$-wave annihilation process, we approximate the thermal averaged cross section 
  in the non-relativistic limit by
\begin{eqnarray}
\langle \sigma v_{rel} \rangle  \simeq \sigma v_{rel} \simeq
\frac{g_{\zeta}^{\, 4}}{16 \pi m_{DM}^{\, 2}} \left(1-\frac{M_{X}^{2}}{m_{DM}^{\, 2}} \right)^{3/2}
\left(1-\frac{M_{X}^{2}}{2 m_{DM}^{\, 2}} \right)^{-2}. 
\end{eqnarray}
With this formula, we solve the Boltzmann equations.  
For the $s$-wave annihilation process, the asymptotic solution of the Boltzmann equation is known, 
  and the relic DM density is approximately given by \cite{Kolb:1990vq, Bertone:2004pz}
\begin{eqnarray}
\Omega_{DM} h^{2}
\simeq \frac{2.13 \times 10^8 \, x_{f}}{\sqrt{g_{\ast}} \, M_{P} \, \langle\sigma v_{rel} \rangle} ,
\end{eqnarray}
where $x_{f}=m_{DM}/T_{f} \simeq \ln(x)-0.5\ln(\ln(x))$ with
 $ x=0.19 \, \sqrt{g_{DM}/g_{\ast}} \, M_{P} \, m_{DM} \, \langle\sigma v_{rel} \rangle $
for the freeze-out temperature $T_{f}$.

%
\begin{figure}[t]
\vspace{-5pt}
\centering
\includegraphics[width=0.55\textwidth]{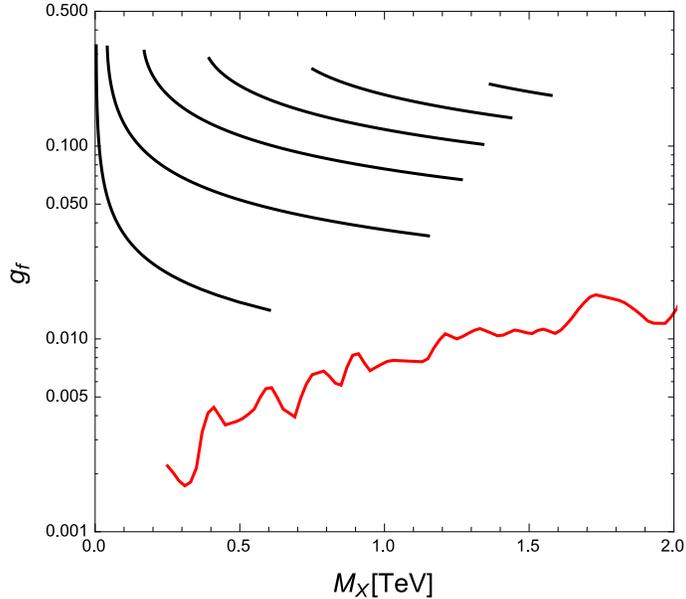} 
\caption{
The plot of $M_X$ versus $g_f$ for the annihilation process (ii) $\bar{\zeta}_{\ell} \, \zeta_{\ell} \rightarrow X \, X$. 
The black solid curves from left to right depict the results for $a=0.3$, $1$, $2$, $3$, $4$ and $5$, 
  respectively, from left to right, along which $\Omega_{DM} h^2=0.12$ is satisfied. 
The diagonal red line shows the upper bound on $g_f$ as a function of $M_X \geq 0.25$ TeV  
  from the LHC Run-2 results. 
No allowed region exists for $M_X \geq0.25$ TeV, which can simultaneously satisfy 
  the cosmological and LHC constraints. 
}
\label{Fig:XX_LHC}
\end{figure}

As an example, we set $m_{DM}= 3 M_X$ in our analysis. 
We find that the results for $m_{DM} > M_X$ is almost independent of $M_X$ 
   unless $m_{DM}$ is taken to be close to $M_X$. 
We have only two free parameters, $g_\zeta$ and $m_{DM}$, 
   involved in this analysis. 
The cosmological constraint to reproduce the observed DM density of $\Omega_{DM} h^2=0.12$ 
   leads to a relation between $g_\zeta$ and $m_{DM}=3M_X$, which is well approximated by
   $g_\zeta \simeq 0.506\, \sqrt{m_{DM}[{\rm TeV}]}$, or equivalently,  
\bea
   M_X[{\rm TeV}] = \frac{1}{3} \left( \frac{g_\zeta}{0.506}\right)^2. 
\eea      
Once $a$ is fixed, $M_X$ is given by a function of $g_\zeta$ in the range of 
  $\left. g_\zeta \right|_{min} \leq g_\zeta \leq \left. g_\zeta \right|_{max}$. 
As in Eq.~(\ref{gfgzeta}), $g_f$ is related to $g_\zeta$ through $g_{BL}$. 
Therefore, $M_X$ is expressed as a function of $g_f$ in the range of 
 $\left. g_f \right|_{min} \leq g_f  \leq \left. g_f \right|_{max}$. 
In Fig.~\ref{Fig:XX_LHC}, we show this relation for $a=0.3$, $1$, $2$, $3$, $4$ and $5$ 
  (black solid curves from left to right), 
  along with the upper bound on $g_f$ from the LHC Run-2 results (diagonal red line). 
We see that for $M_X \geq 0.25$ TeV, the parameter region to reproduce the observed DM density 
  is excluded by the LHC Run-2 result.

Before concluding this section, we comment on the DM physics for $|a| \ll 1$. 
Since $g_\zeta \propto a$, the interaction of the DM particle becomes extremely weak in this case, 
   and the DM particle cannot get in thermal equilibrium with the SM particles. 
In such a case, we consider the so-called freeze-in DM scenario, 
  in which the DM particles are produced from the annihilations of 
  particles in the thermal plasma.  
The analysis of our $X$-portal DM for the freeze-in case is very similar to 
  that in Refs.~\cite{Mohapatra:2019ysk, Okada:2020cue}. 
Following the analysis in these references,   
  we find that the observed DM density is reproduced for $g_\zeta \, g_f \sim 10^{-12}$ and $M_X <  m_{DM}$, 
  independently of $m_{DM}$. 
The condition of  $g_\zeta \, g_f \sim 10^{-12}$ is satisfied by $a \sim 3 \times 10^{-11}$.

\section{Conclusions}
\label{sec7}

Although the (minimal) left-right symmetric extension of the SM (LRSM) based on the gauge group 
  ${\cal G}_{LR}=SU(3)_{c} \times SU(2)_{L} \times SU(2)_{R} \times U(1)_{B-L}$
  is a well-motivated direction to new physics beyond the SM, 
  a candidate of the DM particle in our universe is missing. 
To supplement the LRSM with a suitable DM candidate, we have proposed a minimal extension of the LRSM 
  by introducing a new $U(1)_X$ gauge interaction along with a vector-like fermions $\zeta_{L, R}$
  and a Higgs boson $\phi_X$ which are singlet under $SU(3)_{c} \times SU(2)_{L} \times SU(2)_{R}$. 
Through the spontaneous braking of the gauge symmetry ${\cal G}_{LR} \times U(1)_X$ down to the SM ones, 
  we obtain the extra gauge boson mass eigenstates, $W_R$, $Z_R$ and $X$, 
  and at the same time a Majorana masses for $\zeta_{L, R}$ are generated. 
The lightest Majorana mass eigenstate $\zeta_\ell$ (whose left-handed component is) 
  defined as a liner combination of $\zeta_L$ and ${\zeta_R}^c$
  is stable due to the $U(1)_X$ symmetry and hence the DM candidate in our model. 
For simplicity, we have set the breaking scale of the $SU(2)_R \times U(1)_{B-L}$ to be $v_R=100$ TeV, 
  and focused on $X$-portal DM physics.   
We have considered a variety of phenomenological constraints on this DM scenario 
  to identify the allowed parameter region.

Corresponding to the gauge groups $SU(2)_R$, $U(1)_{B-L}$ and $U(1)_X$,  
   three new gauge couplings, $g_R$, $g_{BL}$ and $g_X$, are involved in our model. 
Imposing the $L \leftrightarrow R$ symmetry, we set $g_R=g$. 
To reproduce the SM hypercharge gauge coupling, $g_{X}$ ($g_{BL}$) is given as 
   a function of $g_{BL}$ ($g_{X}$) and $a$ (the $U(1)_X$ charge of $\phi_X$). 
We have derived the interactions of the $X$ boson with the SM fermions and the Majorana fermion DM $\zeta_\ell$ 
   and obtained the expression of the corresponding gauge couplings $g_f$ and $g_\zeta$ 
   as a function of only two free parameters, $g_{BL}$ and $a$. 
Employing the RG equations at the one-loop level, 
   we have examined the perturbativity condition on the gauge couplings, $g_{BL}, g_X \leq 4 \pi$,
   up to the (reduced) Planck scale and found that the gauge couplings at $v_R$  are constrained 
   to be within certain ranges,  $\left. g_{BL} \right|_{min} \leq g_{BL} \leq  \left. g_{BL} \right|_{max}$
   and $\left. g_{X} \right|_{min} \leq g_{X} \leq  \left. g_{X} \right|_{max}$, once $a$ is fixed. 
The value of $a$ is also constrained to satisfy $|a| \leq 5.28$. 
Correspondingly, $g_f$ and $g_\zeta$ are constrained to be certain ranges for a fixed $a$ values.

If kinematically allowed, the $X$ boson can be produced at the LHC. 
The ATLAS and the CMS collaborations have reported their final LHC Run-2 results
   on the search for a narrow resonance with dilepton final states.  
Calculating the dilepton production cross section through the $X$ boson resonance in our model, 
   we have interpreted the LHC Run-2 results into the upper bound on $g_f$ 
   as a function of $0.25 \leq M_X[{\rm TeV}] \leq 6$. 
Combining this LHC constraint with the result obtained from the perturbativity condition, 
   we have identified a range of $M_X$ for a fixed $a$ value.

Finally, we have investigated the DM physics. 
The DM particle $\zeta_\ell$ communicates with the SM particles (fermions) 
   through the interaction with the $X$ boson. 
In the early universe, the DM particle was in thermal equilibrium with the SM particles, 
   and the DM relic density at present is evaluated by solving the Boltzmann equation. 
We have considered two main processes for the DM pair annihilations: 
  (i) $\zeta_\ell \, \zeta_\ell \to X \to f_{SM} \, \overline{f_{SM}}$ and 
  (ii) $\zeta_\ell \, \zeta_\ell \to X \, X$. 
The process (i) dominates for $m_{DM} \simeq M_X/2$ 
  while the process (ii) dominates for $m_{DM} > M_X$. 
Applying the cosmological constraint, $\Omega_{DM} h^2=0.12$, 
   we have identified the allowed parameter region for the process (i). 
Combining the results with the perturbative condition and the LHC Run-2 constraints, 
     we have found the allowed parameter region to be very narrow. 
As for the process (ii), we have found that the parameter region ($g_f$ as a function of $M_X$) 
   satisfying the cosmological constraint appears far above the upper bound on $g_f$ (for $M_X \geq 0.25$ TeV),
   and no allowed parameter region exists.

 From Fig.~\ref{Fig:DMregions}, we can see that a suitable choice of $m_{DM} \simeq M_X/2$ 
   can reproduce the observed DM density for a wide range of $M_X$ value
   while the severe constraints are from the combination of the LHC results and the perturbativity condition.  
The narrow resonance search at the LHC will continue with the High-Luminosity upgrade of the LHC (HL-LHC). 
In the $X$ boson search with dilepton final states for $M_X > 1$ TeV, 
   the number of the SM background events is very small,
   and we expect that the upper bound on $\sigma(pp\to X \to \ell^+\ell^-)$ will be scaled by $1/{\cal L}$ 
   with the LHC integrated luminosity ${\cal L}$. 
Since $\sigma(pp\to X \to \ell^+\ell^-) \propto g_f^2$ in the narrow decay width approximation, 
    our naive prospect for the HL-LHC experiments with the goal integrated luminosity of ${\cal L}=3000$/fb 
    is that  the current upper bound on $g_f$ shown in the left panel of Fig.~\ref{Fig:LHC}
    will be improved by a factor $\sqrt{139/3000} \sim 0.2$. 
Therefore, a significant portion of the allowed parameter region presented in this paper 
   will be tested at the HL-LHC experiments.

\section*{Acknowledgement}
M.~J.~Neves would like to thanks the Department of Physics \& Astronomy at the University of Alabama
  for the hospitality during his visit as a J-1 Research Scholar.
This work is supported in part by 
  the Conselho Nacional de Desenvolvimento Cient\'ifico e Tecnol\'ogico (CNPq) under grant 313467/2018-8 (GM) (M.~J.~Neves), 
  the United States Department of Energy grant DE-SC0012447 (N.~Okada), and the M.~Hildred Blewett Fellowship
  of the American Physical Society, www.aps.org (S.~Okada).



\end{document}